\documentstyle[supertabular,epsf,rotate]{mn}
\newcommand{\mc}{\multicolumn}
\newcommand{\lta}{\stackrel{<}{_{\sim}}}
\newcommand{\gta}{\stackrel{>}{_{\sim}}}
\newcommand{\ltsimeq}{\raisebox{-0.6ex}{$\,\stackrel 
        {\raisebox{-.2ex}{$\textstyle <$}}{\sim}\,$}} 
\newcommand{\gtsimeq}{\raisebox{-0.6ex}{$\,\stackrel 
        {\raisebox{-.2ex}{$\textstyle >$}}{\sim}\,$}} 

\begin{document}

\title[The emission line -- radio correlation] {The emission line -- radio
correlation for radio sources using the 7C Redshift Survey}

\author[Willott et al.]{Chris J.\ Willott\footnotemark, Steve Rawlings,
Katherine M.\ Blundell and Mark Lacy\\ 
Astrophysics, Department of Physics, Keble Road, Oxford, OX1
3RH, U.K.\\ }

\maketitle

\begin{abstract}

We have used narrow emission line data from the new 7C Redshift Survey
to investigate correlations between the narrow-line luminosities and
the radio properties of radio galaxies and steep-spectrum quasars. The
7C Redshift Survey is a low-frequency (151 MHz) selected sample with a
flux-density limit about 25-times fainter than the 3CRR sample. By
combining these samples, we can for the first time distinguish whether
the correlations present are controlled by 151 MHz radio luminosity
$L_{151}$ or redshift $z$.  We find unequivocal evidence that the
dominant effect is a strong positive correlation between narrow line
luminosity $L_{\mathrm NLR}$ and $L_{151}$, of the form $L_{\mathrm
NLR} \propto L_{151}^{0.79 \pm 0.04}$.  Correlations of $L_{\mathrm
NLR}$ with redshift or radio properties, such as linear size or 151
MHz (rest-frame) spectral index, are either much weaker or absent. We
use simple assumptions to estimate the total bulk kinetic power $Q$ of
the jets in FRII radio sources, and confirm the underlying
proportionality between jet power and narrow line luminosity first
discussed by Rawlings \& Saunders (1991). We make the assumption that
the main energy input to the narrow line region is photoionisation by
the quasar accretion disc, and relate $Q$ to the disc luminosity,
$Q_{\mathrm phot}$.  We find that $0.05 \ltsimeq Q/ Q_{\mathrm phot}
\ltsimeq 1$ so that the jet power is within about an order of
magnitude of the accretion disc luminosity. Values of $Q/ Q_{\mathrm
phot} \sim 1$ require the volume filling factor $\eta$ of the
synchrotron-emitting material to be of order unity, and in addition
require one or more of the following: (i) an important contribution to
the energy budget from protons; (ii) a large reservoir of
mildly-relativistic electrons; and (iii) a substantial departure from
the minimum energy condition in the lobe material.  The most powerful
radio sources are accreting at rates close to the Eddington limit of
supermassive black holes ($M_{\mathrm BH} \gtsimeq 10^{9} M_{\odot}$),
whilst lower power sources are accreting at sub-Eddington rates.

\end{abstract}

\begin{keywords}
galaxies:$\>$active -- galaxies:$\>$jets -- quasars:$\>$general 
-- quasars:$\>$emission lines
\end{keywords}

\footnotetext{Email: cjw@ll.iac.es \\ Present address: Instituto de
Astrof\'\i sica de Canarias, C/ Via Lactea s/n, 38200 La Laguna,
Tenerife, Spain}

\section{Introduction}
\label{sec:intro}

Despite all the observational effort which has gone into researching
extragalactic radio sources, we are still uncertain as to how these
objects are powered and the relationship between the active nuclei,
host galaxies and larger-scale environments. The current favoured
models of powering extragalactic radio sources involve gas accretion
onto a supermassive black hole ($M_{\mathrm BH} \gg 10^{6}
M_{\odot}$), but details more precise than this remain controversial.
 
There is a strong positive correlation between the extended radio
luminosities and narrow emission line luminosities of 3C radio sources
(Baum \& Heckman 1989; Rawlings et al. 1989; McCarthy 1993;
Tadhunter et al.\ 1998). Rawlings \& Saunders (1991, hereafter RS91) 
argued that by making an estimate of the bulk kinetic power in the 
jet from the observed low radio-frequency luminosity, it was
possible to tighten this correlation (see also Falcke, Malkan \&
Biermann 1995).  This implies that the energy source of the
narrow lines is inextricably linked to the source of the radio emission. The
accretion rate and/or black hole mass of the central engine are
possible candidates for the connection between both types of emission.
A correlation between the optical and extended-radio luminosities of
steep-spectrum quasars has recently been confirmed (Serjeant et
al. 1998; Willott et al. 1998a), and appears to argue strongly against
alternative models in which radio and narrow-line luminosities are
controlled mainly by the environment, rather than the central engine
(e.g. Dunlop \& Peacock 1993).

There is considerable and mounting evidence that radio galaxies and
radio-loud quasars are the same objects viewed at different angles to
the radio jet axis (Scheuer 1987; Barthel 1989; Antonucci 1993). A
dusty torus perpendicular to the jet axis, with a high optical depth,
obscures the central emission regions if the jet is roughly in the
plane of the sky and a radio galaxy is observed.  For objects where
our line-of-sight is within the radiation cone defined by the torus, we
see a luminous continuum source and broad emission lines, and the
object is called a quasar. This simple model of unification by
orientation appears to work remarkably well. Scattered broad emission
lines have been observed in several radio galaxies (e.g.\ 3C~234,
Antonucci 1984; 3C~324, Cimatti et al. 1996; 3C~265, Dey \& Spinrad
1996; Cygnus A, Ogle et al. 1997), in accordance with the unified
schemes.  Near- and thermal-infrared studies (e.g.\ Hill, Goodrich \&
DePoy 1996; Simpson, Rawlings \& Lacy 1999) of radio sources have
revealed quasar nuclei obscured by rest-frame visual extinctions
$A_{\mathrm V}$ in the range $0 < A_{\mathrm V} < 15$.

The narrow-line region (NLR) in radio sources is extended over several
kpc, beyond the putative dusty torus, and therefore the narrow-line
emission is believed to be (largely, and perhaps entirely) independent
of the jet axis orientation.  Similar distributions of narrow-line
luminosities in samples of radio galaxies and quasars, matched in
extended radio luminosity, have been suggested as a fundamental test
of the unified schemes (e.g.\ Barthel 1989). At low redshift ($z<0.8$),
there have been claims that quasars are observed to have [OIII] line
luminosities a factor of 5-10 greater than radio galaxies of similar
radio luminosities (Baum \& Heckman 1989; Jackson \& Browne 1990;
Lawrence 1991). However, the [OII] luminosities of radio galaxies and
quasars at low redshift are indistinguishable (Browne \& Jackson 1992;
Hes et al.\ 1993). This has been explained as partial obscuration of
[OIII] emission in radio galaxies, because [OIII] is emitted from a
region closer to the nucleus than [OII], due to its higher ionisation
state. However, Jackson \& Rawlings (1997; JR97) have investigated the
[OIII] luminosities of $z>1$ radio galaxies and quasars and find their
distributions indistinguishable. A possible resolution of these
perceived difficulties with the unified schemes has recently been
given by Simpson (1998).

Much has been learnt about radio sources from the bright 3CRR radio
catalogue (Laing, Riley \& Longair 1983 - LRL, see Section 2.2).
However, due to the tight luminosity--redshift correlation present in
a single flux-limited sample, it is impossible to distinguish whether
the differences observed between 3CRR objects at high and low
redshifts are correlations with redshift, or a consequence of the
different radio luminosities of the sources.  The best way to
disentangle luminosity and redshift effects is to study a fainter
radio sample and combine it with the 3CRR sample. To minimise
selection effects the selection frequency of the faint sample should
be close to or slightly lower than that of 3CRR (178 MHz). As
discussed in detail by Blundell, Rawlings \& Willott (1999a, hereafter
BRW99), samples selected at significantly higher frequencies have
significantly different features: namely, (i) preferential selection
of Doppler-boosted objects whose radio axes are not randomly oriented
with respect to the line-of-sight; and (ii) decreased sensitivity to
orientation-independent steep-spectrum lobe emission.  The latest
published work on the radio--optical correlation in a large complete
sample of radio sources (Tadhunter et al. 1998) has neither the
benefits of going to fainter radio fluxes than 3CRR (for
steep-spectrum objects), nor those of selecting at low radio
frequencies.  Allington-Smith, Peacock \& Dunlop (1991) have made a
limited study of emission line strengths in a 408 MHz-selected sample
containing, at a given redshift, steep-spectrum objects which are
slightly less radio-luminous than 3CRR objects.

We have measured the redshifts and emission line fluxes of $>90\%$ of
the 77 radio sources in two of the three regions of sky which comprise
the 7C Redshift Survey. A brief description of this
new survey is given in Section 2.1. At a given redshift, 7C sources 
are a factor of $\sim 25$ times lower in luminosity than 3CRR sources. 
Hence, we can now for the first time clearly distinguish between redshift- and
luminosity-dependent effects. In this paper we use the emission line
luminosities of the 7C sources, in combination with 3CRR data from the
literature, to investigate correlations between radio and
emission-line properties of radio galaxies and quasars. The convention
for radio spectral index, $\alpha$, is that $S_{\nu} \propto
\nu^{-\alpha}$, where $S_{\nu}$ is the flux density at
frequency $\nu$. We assume throughout that $H_{\circ}=50~ {\rm
km~s^{-1}Mpc^{-1}}$, $q_{0}=0.5$. The quantitative results are
slightly different if one assumes $q_{0}=0$, as discussed in the
text.

\section{The complete samples}

\subsection{7C Redshift Survey}

Only brief details of the 7C Redshift Survey are given here; full
details will appear elsewhere (see also Rawlings et al. 1998, Willott
et al. 1998a, 1998b, BRW99, Blundell et al. 1999b).  The 7C-I and
7C-II regions contain a total of 77 radio sources which have all been
identified with an optical/near-IR counterpart. These samples were
selected to include all sources with flux-densities $S_{151}\geq 0.5$
Jy at 151 MHz over 0.013 sr.  Spectroscopic redshifts have been
obtained for all but 6 of these sources (Willott et al., in
prep.). For these 6 objects, which are expected to be at $z>1$ from
their $K$-band magnitudes, we have undertaken near-infrared photometry
and spectroscopy in order to constrain their redshifts (Rawlings et
al., in prep.). For 5 of these 6 objects, their spectral energy
distributions (SEDs) are well-fitted by model elliptical galaxies with
redshifts in the range $1<z<2$. The other object (5C7.47) has an
unusual SED, and its redshift is firmly constrained only by the lack
of a Lyman-limit cutoff in its spectrum, yielding $z < 4$, but for
reasons discussed in Section 2.4 we take $z = 1.5$.
As a result of the low selection frequency of the 7C sample, only one
object (5C7.230) lies above the sample flux limit only because its
core flux is Doppler boosted; this object is excluded from the
analysis presented in this paper. One of the 7C sources is also in the
3CRR sample (3C 200), so excluding this too, the final 7C sample
comprises 75 sources.

\subsection{3CRR sample}

The bright radio sample used is the 3CRR sample of LRL, which has
complete redshift information for all 173 sources, selected with
$S_{178}\geq 10.9$ Jy. 3C 345 and 3C 454.3 are flat-spectrum quasars
which are excluded on the grounds of Doppler-boosting raising their
fluxes above the selection limit. 3C 231 (M82) is excluded because it
is a very nearby starburst galaxy and not a radio-loud AGN. Hence our
revised 3CRR sample consists of 170 sources. 
 
\subsection{Optical classification of radio-loud AGN}

Historically, quasars were defined as luminous objects with unresolved
`stellar' IDs and strong, broad (FWHM $> 2000$ km s$^{-1}$) emission
lines. In contrast, radio galaxies have a resolved optical appearance
and narrow (FWHM $\approx 500$ km s$^{-1}$) or absent emission lines.
However, it is now apparent that this simple classification scheme is
inadequate: for example, weak quasars with host galaxies visible may
be classified as broad-line radio galaxies (BLRGs). As mentioned in
the introduction, spectropolarimetry of some radio galaxies has
revealed broad lines scattered into our line-of-sight. It is also
clear that some quasars have been reddened by dust so that the quasar
is significantly dimmed in the UV (e.g. 3C 22: Rawlings et al. 1995;
3C 41: Simpson et al. 1999).  Therefore the simple classification that
quasars have broad emission lines must be treated with caution
because, even with small amounts of reddening, this becomes dependent
on redshift and whether near-infrared data is available.

For these reasons we define quasars as objects with an unresolved
nuclear source which itself has absolute magnitude $M_{B}<-23$
($q_{0}=0.5$) after eliminating light from any host galaxy and/or
other diffuse continuum (such as radio-aligned emission, e.g. McCarthy
et al. 1987). Some objects with broad lines have nuclei fainter than
this limit. These objects will be defined in this paper simply as weak
quasars (WQs). The weak quasars can be split further into three groups
with sufficient polarimetric/near-infrared data; dust-reddened quasars
with $A_{\mathrm V} \approx 1-2$, traditional broad-line radio
galaxies which have intrinsically weak AGN, and radio galaxies with
scattered broad lines. Two 3CRR sources have been reclassified since
the classifications given in Willott et al. (1998a). 3C 318 has been
found to be a quasar at $z=1.574$, not a broad-line radio galaxy at
$z=0.752$ (Willott, Rawlings \& Jarvis 1999). 3C 343 does not have
broad emission lines in the spectrum of Lawrence et al. (1996) and is
clearly resolved in the HST imaging of Lehnert et al. (1999).
Therefore it is classified as a radio galaxy, not a quasar.

There are 23 quasars and 2 WQs in the 7C sample. The 3CRR sample
contains 40 quasars and 13 WQs. Therefore, there are a total of 78
FRII quasars/WQs in our combined sample\footnotemark. There are 50
radio galaxies in the 7C sample of which only 5 have FRI structures,
the rest being FRIIs or possible FRIIs (objects with a defined angular
size; DAs in the notation of BRW99). To avoid confusing the issue we
will hereafter refer to all objects which are not FRIs as FRIIs.  In
3CRR there are 117 radio galaxies (93 FRIIs and 24 FRIs).

\footnotetext{Note that we have slightly adapted the original Fanaroff
\& Riley (1974) scheme for classifying the structure of radio
sources. Details of this adaptation can be found in BRW99, and full
lists of structural classification in Blundell et al. 1999b. Although
the radio structures of the quasars 5C6.264 and 3C 48 appear to be
FRIs, in this paper they are deemed to be FRIIs given their high radio
luminosities.}

\subsection{Narrow-line luminosities}

The 245 radio sources in our combined (7C + 3CRR) sample have
redshifts in the range $0 < z < 3$, so it is not possible to measure
the flux of the same emission line in each object via optical
spectroscopy. The [OII] $\lambda 3727$ line has been used for all
sources where it was observed, since it is the most frequently
observed narrow line in the quasars and radio galaxies in the combined
sample. In cases where [OII] was not observed, other narrow lines such
as [OIII] $\lambda 5007$, H$\alpha$ (RGs only), and Ly$\alpha$ (RGs
only) were measured instead and converted to the expected [OII] flux
using the average line ratios quoted by McCarthy (1993).  Note that
this method is imperfect because there exists significant scatter in
the line ratios measured for radio galaxies, and evidence for some
systematic changes in excitation-sensitive line ratios (like
[OII]/[OIII]) with radio luminosity (Saunders et al. 1989; Tadhunter
et al.  1998); it is also possible that the McCarthy line ratios are
biased towards exceptionally radio luminous objects over some
spectral, and hence redshift, ranges.  Line fluxes and redshifts for
7C quasars and weak quasars are in Willott et al. (1998a). The spectra
and line measurements of the 7C radio galaxies will be presented
elsewhere (Willott et al., in prep.). The line luminosities for 3CRR
sources were taken primarily from the compendia of JR97 and Hirst,
Jackson \& Rawlings (1999), with additional data from R. Laing
(priv. comm.) and J. Wall (priv. comm.). The complete updated list of
3CRR line luminosities and their references are available on the
worldwide-web at {\bf http://www.iac.es/galeria/cjw/3crr.html}.

\begin{figure*}
\epsfxsize=0.9\textwidth
%\begin{flushleft}
%\hspace{-1.3cm}
%\hspace{0.4cm}
\epsfbox{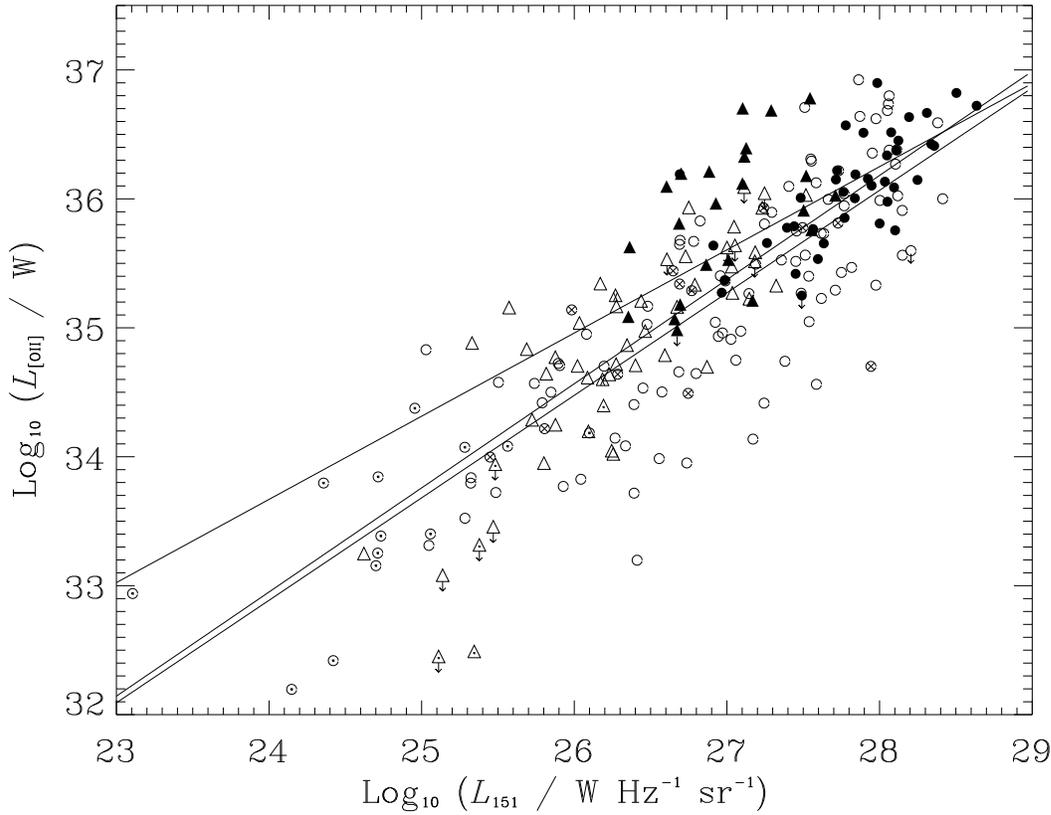}
%\end{flushleft}
{\caption[junk]{\label{fig:loiilr}
[OII] emission line luminosity, $L_{\mathrm [OII]}$, against
(rest-frame) 151 MHz radio luminosity $L_{151}$ for the 7C
(triangles) and 3CRR (circles) complete samples. The radio flux
densities and spectral indices needed to calculate $L_{151}$ are
tabulated in Blundell et al. (1999b). The sources from each sample are
classified into four groups: quasars (solid symbols), weak quasars
(open with a cross), FRII radio galaxies (open) and FRI radio galaxies
(open with a dot). The solid lines show the power-law fits to (from top
at left) (i) all the quasars and weak quasars, (ii) all sources
classified as FRII, (iii) all sources classified as FRII radio
galaxies.}}
\end{figure*}

FRI radio sources are known to have a different radio-optical
correlation from FRIIs (Zirbel \& Baum 1995). Of the 24 FRI radio
galaxies in 3CRR, there are no emission line data in the literature
for 11 of them. Therefore, we can only calculate line luminosities for
about half of the 28 FRI galaxies in the combined sample. This is
probably a biased subset of the true FRI population because
objects with brighter lines are more likely to have had line fluxes
published.  Due to this incompleteness, we do not attempt to use the
observed sources to investigate the FRI radio--optical correlation.

For quasars at redshifts $z\gtsimeq 1.3$, the [OII] line passes out of
the wavelength region of our optical spectra. There are few other
prominent narrow quasar emission lines below the wavelength of [OII],
so for most of the high-$z$ quasars, the only narrow line fluxes
measured are those from the near-infrared spectrophotometry of JR97,
Hirst et al. (1999) and Willott et al. (in prep.). For other
high-redshift quasars the [OII] flux was estimated by calculating the
continuum level at the wavelength of [OII] from the magnitudes in
Willott et al. (1998a) and LRL (for 7C and 3CRR, respectively) and
assuming that the rest-frame equivalent width of the [OII] line is
10\AA~, a typical value for `naked' quasars (Miller et al. 1992).  A
total of 12 out of the 78 quasars in the combined sample have [OII]
fluxes estimated in this way. They have a similar distribution of
$L_{\mathrm [OII]}$ to the quasars with measured narrow lines. Use of
this method, which again introduces some scatter because of the finite
spread in [OII] equivalent widths for quasars, means we can include
all the quasars and weak quasars in our analysis.

There are no line fluxes (or limits) in the literature for 5 of the
3CRR FRII radio galaxies.  This is such a small fraction of the total
number of sources that their exclusion should not significantly affect
our results. 

For 8 FRII radio galaxies in 7C and 2 FRIIs in 3CRR, only upper limits
to the emission line luminosities are known. Where there is only an
upper limit available for a line luminosity, we assume that the line
luminosity is equal to this limit. One quasar from each sample has a
line limit fainter than that calculated assuming an equivalent width
of 10 \AA, so the limit is adopted in these cases as well. There are 6
FRII radio galaxies in the 7C sample which do not have spectroscopic
redshifts. For the 5 objects with redshifts well-constrained by SED
fitting, we can set a limit on their H$\alpha$ fluxes from our near-IR
spectra.  These objects have estimated redshifts which are in the
range which is traditionally difficult to measure, because no strong
lines (e.g. Ly$\alpha$, [OII]) appear in the observed optical
spectrum.  For these objects the H$\alpha$ limits from near-infrared
spectroscopy lie within the spread of the $L_{\mathrm [OII]}-L_{151}$
correlation, so there is no reason to expect them to have
significantly weaker lines than other sources, their lack of observed
emission lines being a simple consequence of their redshifts. For the
object 5C7.47, we assume its redshift lies in the difficult range
($1.3 \leq z \leq 1.8$), and set a limit on its line strength from our
optical spectrum, associating this limit with the MgII $\lambda 2798$
line.

\begin{table*}
\footnotesize
\begin{center}
\begin{tabular}{lrlccr}
\hline\hline 
\mc{1}{l}{Sample}& \mc{1}{c}{N}& \mc{1}{l}{Correlated}&
\mc{1}{c}{r$_{\mathrm AB}$}& \mc{1}{c}{r$_{\mathrm AB,C}$}&
\mc{1}{c}{signif} \\
\mc{1}{l}{} & \mc{1}{l}{} & \mc{1}{l}{variables: A,B} & \mc{1}{c}{} &
\mc{1}{c}{} & \mc{1}{c}{$\sigma$} \\
All FRIIs & 211 & $L_{151},L_{\mathrm [OII]}$& 0.784 & 0.635 & 10.78 \\
All FRIIs & 211 & $z,L_{\mathrm [OII]}$ & 0.679 & 0.405 & 6.18 \\
All FRIIs & 211 & $L_{151},z$ & 0.613 & 0.176 & 2.56 \\
Quasars \& WQs& 78 & $L_{151},L_{\mathrm [OII]}$& 0.651 & 0.579 & 5.68 \\
Quasars \& WQs& 78 & $z,L_{\mathrm [OII]}$ & 0.539 & 0.427 & 3.92 \\
Quasars \& WQs& 78 & $L_{151},z$ & 0.366 & 0.023 & 0.19\\
FRII RGs & 133 & $L_{151},L_{\mathrm [OII]}$& 0.794 & 0.617 & 8.18 \\
FRII RGs & 133 & $z,L_{\mathrm [OII]}$ & 0.699 & 0.376 & 4.49 \\
FRII RGs & 133 & $L_{151},z$ & 0.665 & 0.253 & 2.94 \\
\hline\hline
\end{tabular}
\end{center}              
{\caption[Table of correlations]{\label{tab:table} Spearman partial
rank correlation analysis of the correlations present between
$L_{151},L_{\mathrm [OII]}$ and $z$ for various subsets of the
data. This method (e.g. Macklin 1982) assesses the statistical
significance of correlations between the two named variables in the
presence of the third. r$_{\mathrm AB}$ is the rank correlation
coefficient of the two variables and r$_{\mathrm AB,C}$ the partial
rank correlation coefficient. The significance of the partial rank
correlation is equivalent to the deviation from a unit variance normal
distribution if there is no correlation present.  }}

\normalsize
\end{table*}

\begin{figure*}
\epsfxsize=0.9\textwidth
%\begin{flushleft}
%\hspace{-1.3cm}
%\hspace{0.4cm}
\epsfbox{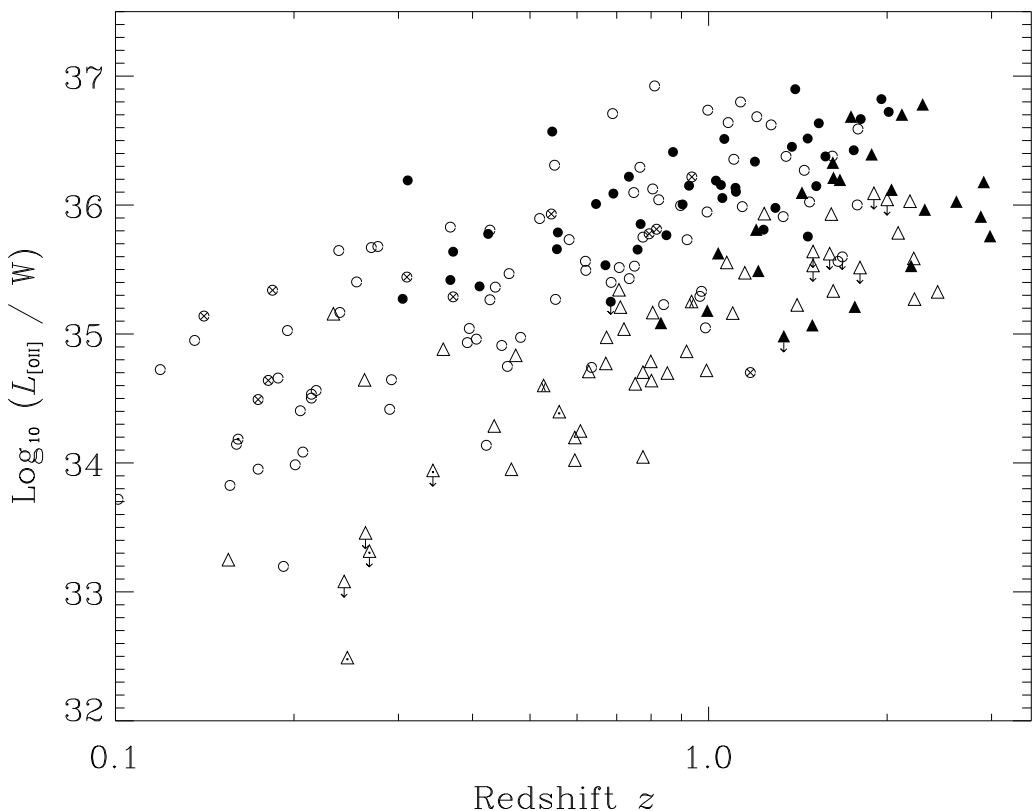}
%\end{flushleft}
{\caption[junk]{\label{fig:loiiz} $L_{\mathrm [OII]}$ versus $z$ for
all the $z>0.1$ sources only. Symbols as for Fig. \ref{fig:loiilr}.
Note the vertical separation between 3CRR (circles) and 7C (triangles)
sources, which is direct evidence that $L_{\mathrm [OII]}$ correlates
primarily with $L_{151}$.}}
\end{figure*}

In conclusion, we have line luminosities or limits for 211 FRII
sources out of a total of 216 FRIIs in 7C and 3CRR. We do not expect
to be significantly biased in any way by the exclusion of the missing
few sources, or by the 5\% of cases where we have only upper limits.  Using
statistical techniques which account for upper limits make negligible
differences to the results of this paper.

\begin{figure*}
\epsfxsize=0.9\textwidth
%\begin{flushleft}
%\hspace{1.0cm}
%\hspace{0.4cm}
\epsfbox{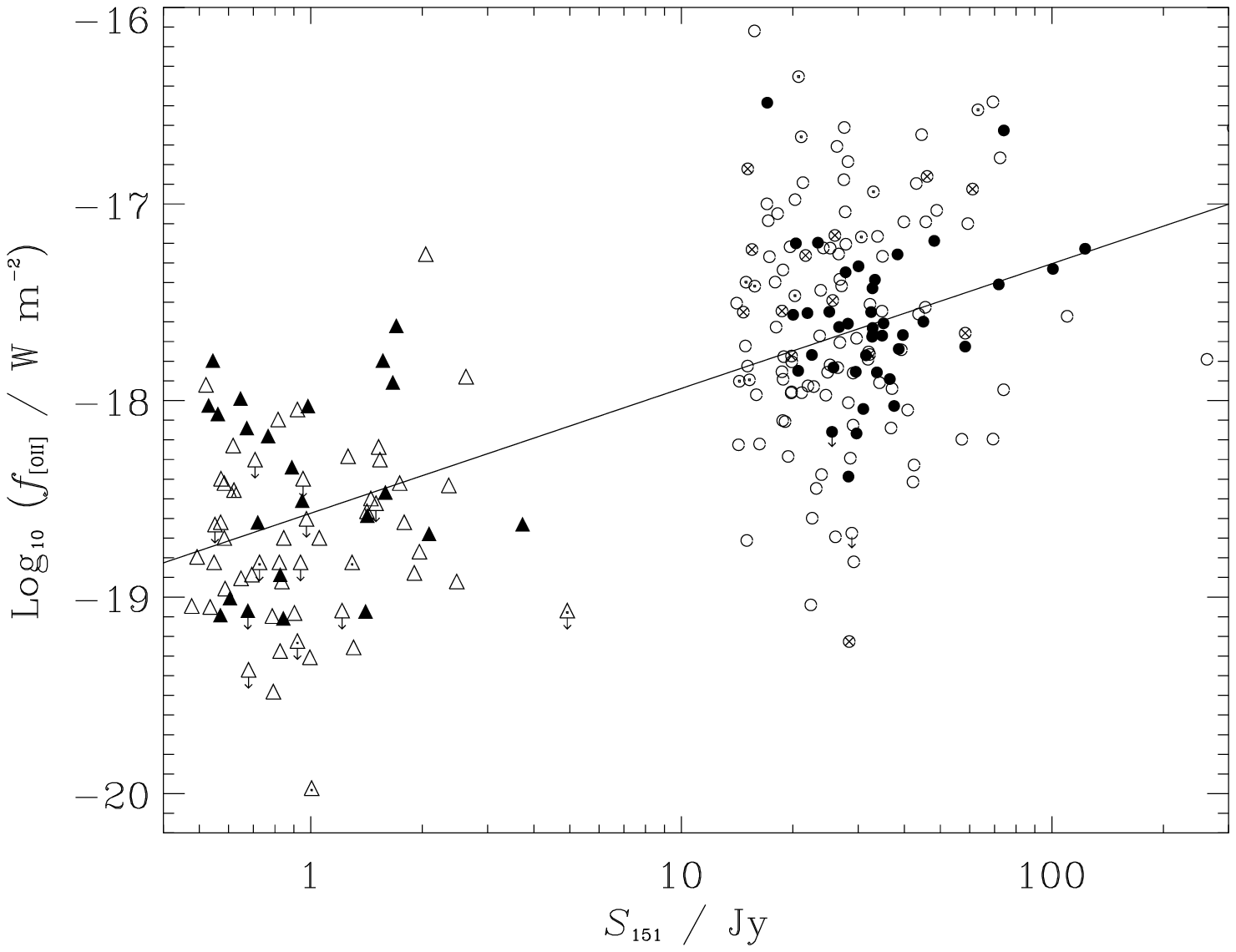}
%\end{flushleft}
{\caption[junk]{\label{fig:foiifr}Emission line flux against radio
flux-density for sources in the combined 7C+3CRR sample. All 229
sources are shown except 3C 274 (also known as the FRI radio galaxy M87 -
excluded from this plot because its radio and [OII] fluxes are much
greater than any other sources in the sample). Symbols are as for
Fig. \ref{fig:loiilr}. The solid line goes through the median emission
line and radio fluxes for each of the two samples.  }}
\end{figure*}

\section{Correlation of narrow-line and radio properties}

\subsection{$L_{\mathrm [OII]}$ versus radio luminosity, $L_{151}$} 

In Figure \ref{fig:loiilr} we plot the [OII] line luminosity against
low-frequency radio luminosity for the sample of 211 FRII sources (the
18 FRIs for which we have emission line data are also plotted). The
strong correlation discussed in Section 1 is clearly apparent. The
line luminosities of the 3CRR and 7C sources are about the same at
radio luminosities at which the samples overlap, despite their vastly
different redshifts. The best-fit power-law solutions for this
correlation are plotted for several subsets of the sample (calculated
by minimising the sum of the squares of the residuals). The slope of
the relation for all FRII sources is $0.81 \pm 0.04$ ($0.85 \pm 0.04$
for $q_{0}=0$), whilst that for just the quasars and weak quasars is
$0.65 \pm 0.07$ ($0.72 \pm 0.07$ for $q_{0}=0$). We discuss
this marginally significant difference between radio galaxies and
quasars in another paper (Willott et al. 1999), noting here the
possibility that it is at least partly an artefact of the way in which
quasars and radio galaxies are discriminated (Section 2.3; see also
Simpson 1998). There is considerable scatter about the best-fit
correlation of Fig. \ref{fig:loiilr}. This scatter is approximately
gaussian in terms of $\log_{10} L_{\mathrm [OII]}$, with a standard
deviation of 0.5.

Figure \ref{fig:loiiz} plots $L_{\mathrm [OII]}$ against redshift for
the same data. There is a clear separation between the 3CRR and 7C
points, with each sample showing its own $L_{\mathrm [OII]}-z$
correlation which Fig. \ref{fig:loiilr} shows is primarily due to
the $L_{\mathrm [OII]}-L_{151}$ correlation rather than to an inherent
$L_{\mathrm [OII]}-z$ correlation for radio sources.  The Spearman
partial rank correlation method (Macklin 1982) was used to check this
result, and Table 1 shows the results of this analysis. For the
combined sample, the correlation between line luminosity and radio
luminosity is much stronger than the correlation of either property
with $z$. However, note that there are still significant correlations
between $L_{\mathrm [OII]}$ and $z$, independent of $L_{151}$, and
$L_{151}$ and $z$, independent of $L_{\mathrm [OII]}$. The correlation
between $L_{151}$ and $z$ is a selection effect, caused by using
flux-limited samples.

The emission line flux -- radio flux-density relation for the 3CRR and
7C complete samples is shown in Figure \ref{fig:foiifr}. It is clear
that the brighter radio sources also have brighter emission lines. The
median [OII] flux for 3CRR FRII sources is $2.2 \times 10^{-18}$
Wm$^{-2}$. The median for 7C FRII sources (which are approximately 25
times fainter in radio flux-density than 3CRR) is $2.4 \times
10^{-19}$ Wm$^{-2}$. The ratio of the median [OII] flux and the median
radio flux density for the two samples implies a slope of 0.63, as
plotted in Fig. \ref{fig:foiifr}. Given the broadly similar redshift
distributions of the 3CRR and 7C samples, this is incontrovertible
evidence that $L_{\mathrm [OII]}$ is predominantly correlated with
radio luminosity, not redshift.

\subsection{$L_{\mathrm [OII]}$ versus linear size, $D$} 

\begin{figure*}
\epsfxsize=0.9\textwidth
%\begin{flushleft}
%\hspace{-1.3cm}
%\hspace{0.4cm}
\epsfbox{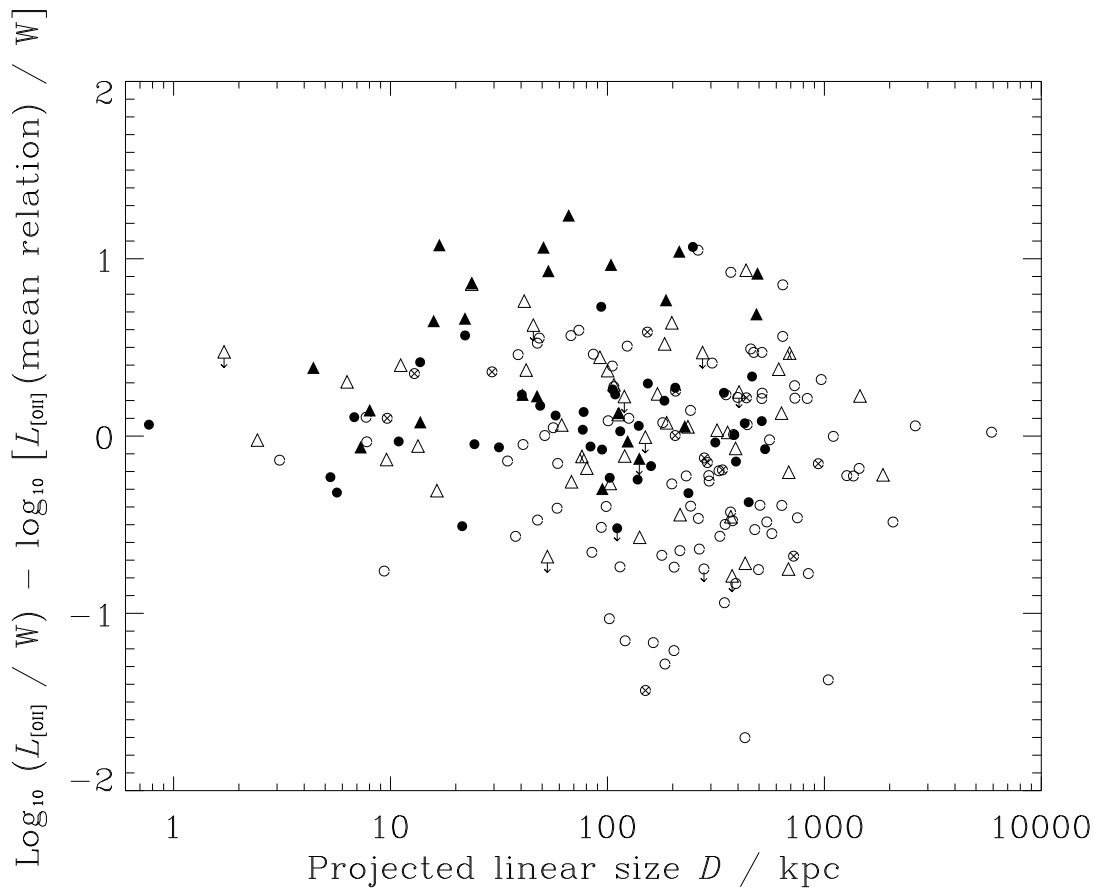}
%\end{flushleft}
{\caption[junk]{\label{fig:loiid} Residual emission-line luminosity
(after subtraction of best-fit $L_{\mathrm [OII]} - L_{151}$
power-law) against projected linear size. Only FRII radio sources are
plotted here. Symbols are as for Fig. \ref{fig:loiilr}.}}
\end{figure*}

We next investigate whether there is any dependence of the line
luminosity on the size of the radio source, independent of the
correlation with low-frequency radio luminosity. In Figure
\ref{fig:loiid}, we have subtracted the power-law fit of $L_{\mathrm
[OII]} - L_{151}$ for all FRII sources from their [OII] luminosities
and plot the residual against projected linear size. There is a weak
anti-correlation here ($r_{DL_{\mathrm [OII]res}}=-0.18$, with
significance 99.1\%). A power-law fit gives a slope of $-0.14 \pm
0.05$. Possible causes of this weak correlation and the large scatter
will be discussed in Section 4.3. It is interesting to note that
virtually all the relatively weak lined objects [$\Delta \log_{10}
(L_{\mathrm [OII]})<-0.5$] have $D \gta 100$ kpc.

\subsection{$L_{\mathrm [OII]}$ versus low-frequency radio spectral index, $\alpha_{151}$} 

\begin{figure*}
\epsfxsize=0.9\textwidth
%\begin{flushleft}
%\hspace{-1.3cm}
%\hspace{0.4cm}
\epsfbox{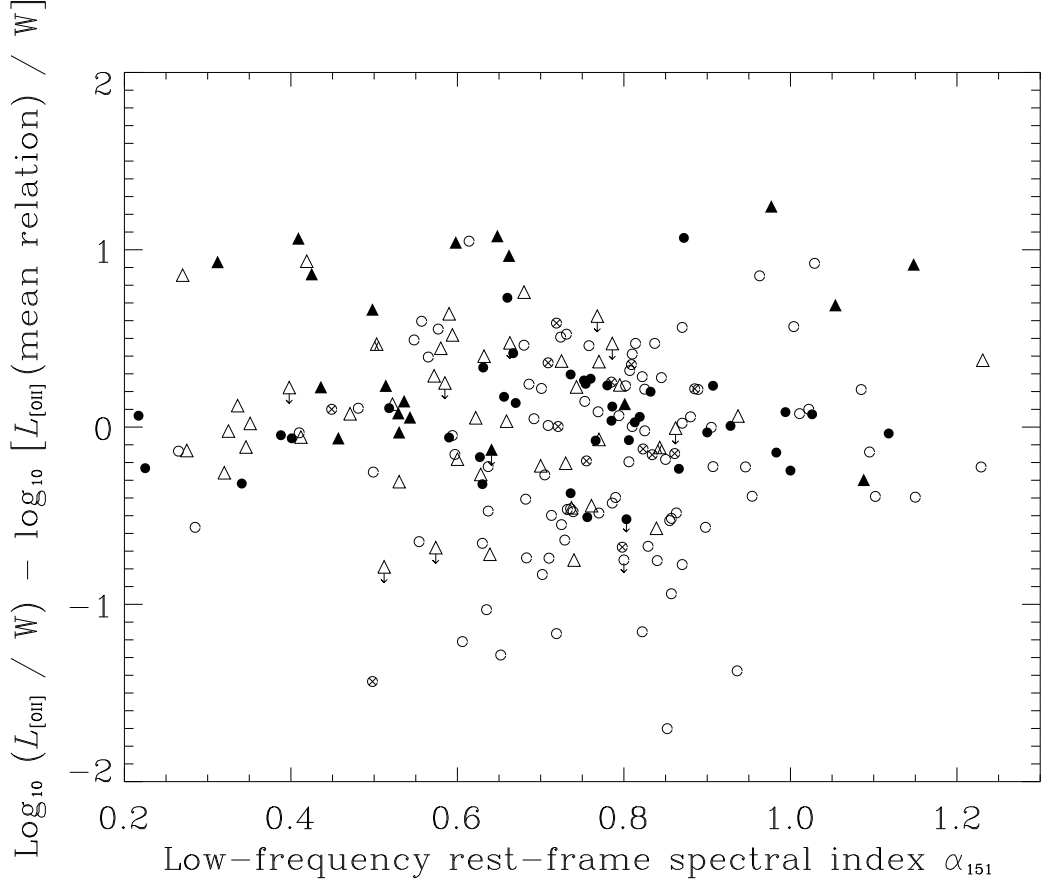}
%\end{flushleft}
{\caption[junk]{\label{fig:loiialp} Residual emission-line luminosity
against rest-frame low-frequency (151 MHz) spectral index,
$\alpha_{151}$. Only FRII sources are plotted. Symbols are as for
Fig. \ref{fig:loiilr}. The two sources with negative spectral indices
(5C6.291 and 3C286) are not plotted for clarity.}}
\end{figure*}

Figure \ref{fig:loiialp} plots the residual [OII] luminosity (after
subtraction of the line--radio luminosity correlation, as in Sec. 3.2)
against the rest-frame 151 MHz spectral index, $\alpha_{151}$, for all
FRIIs in the sample. The Spearman rank correlation co-efficient is
$r_{\alpha_{151} L_{\mathrm [OII]res}}=-0.105$, with significance
87\%. Hence we find no evidence for a correlation between spectral
index and emission line luminosity, independent of radio luminosity.

\section{Discussion}

\subsection{The physical meaning of extended radio luminosity}

\subsubsection{Minimum energy formalism}

For most FRII radio sources only a small fraction of the total kinetic
power of the jets is released as synchrotron radiation in the lobes and
hotspots. A much greater fraction is stored in the lobes and/or lost to the
environment via work done by the expanding radio source (Scheuer 1974). 
We will define the time-averaged jet power $\bar{Q}$ as the total energy 
transported from the central engine by both jets 
divided by the age of the radio source. 
We reserve the symbol $Q_{o}$ to refer to the
instantaneous power transported by one jet, 
and the symbol $Q = 2 Q_{o}$ for the instantaneous power
of both jets, noting that these quantities might vary with time.
An estimate of $\bar{Q}$ can be obtained from the minimum stored energy
required in the lobes to produce the observed synchrotron luminosity, 
the age of the radio source, and consideration of the efficiency with which
$Q$ is converted into the internal energy of the 
observable synchrotron-emitting population.

We first review the calculation of the minimum energy density
$u_{\mathrm min}$ in the lobes of a radio source (see also Leahy 1991). 
We will introduce various factors (denoted by $f$ and $g$)  
to allow for systematic uncertainties,
and define these so that their values lie above unity.  
Cylindrical symmetry is assumed so the source has a volume, 
$V=\pi (D/2 R_{T})^{2}D/\sin\theta$, where: $D$
is the projected linear size of the source; $R_{T}$ is the axial
ratio, the ratio of the total measured length to the
total measured width as defined by Leahy \& Williams (1984)
and Leahy, Muxlow \& Stephens (1989); and
$\theta$ is the angle the jet-axis makes with
the line of sight. In subsequent calculations
we make the approximation that $R_T=5$, and that 
$\theta = 60^{\circ}$, the appropriate average value for 
a randomly distributed sample of radio sources; source-to-source
variations in $R_T$ and $\theta$ will produce some scatter, 
but negligible systematic effects.
The minimum energy density is given by (e.g. Miley 1980)
\begin{displaymath}
u_{\mathrm min}=3.10^{-11} ~\times
\end{displaymath}
\begin{equation}
\label{eq:miley}
\left[ 
\frac { S (1+k) \nu^{\alpha} \left(
\nu_{\mathrm low}^{\frac{1}{2}-\alpha} - \nu_{\mathrm
high}^{\frac{1}{2}-\alpha} \right) (1+z)^{3+\alpha} } 
{\sigma \eta L
\left( \alpha - \frac{1}{2} \right) 
\sin^{3/2} \phi} \right] ^{4/7} {\mathrm J \, m^{-3}}, 
\end{equation}
where: $S/\sigma$ is the source surface brightness in Jy arcsec$^{-2}$;
$k$ is the reciprocal of the ratio of the energy in 
synchrotron-radiating particles (i.e. relativistic electrons and/or positrons) 
to the energy in other particles (e.g. hot and/or cold
protons); $\nu$, the observing frequency in GHz; 
$\alpha$, the radio spectral index;
$z$, the redshift of the source; $\phi$, the angle between the
magnetic field direction and the line-of-sight, which we set to
$90^{\circ}$, and introduce a factor $f_{\rm geom}$ to account 
for systematically-lower values; $\eta$ is the volume 
filling factor; $L$ is the line-of-sight depth through the source 
in kpc which in our assumed geometry equals $D/(R_{T} \sin 60^{\circ})$;
and $\nu_{\mathrm high}$ and $\nu_{\mathrm low}$ 
are the upper and lower cut-off
frequencies of the synchrotron spectrum. 
For $\alpha > 0.5$
in the radio lobes, inspection of eqn.~\ref{eq:miley} shows that 
$u_{\mathrm min}$ is fairly insensitive to the 
upper cut-off frequency (we take 100 GHz), 
but does depends critically on the lower frequency cut-off. The true value
of $\nu_{\mathrm low}$ for FRII sources is unknown because
observations are not possible from the Earth for $\nu\ltsimeq10$
MHz. To account for this uncertainty, which is ultimately due to 
our ignorance of the energy density of mildly relativistic particles, 
we adopt a cut-off at a rest-frame frequency 10 MHz, and introduce the 
factor $f_{\mathrm lowE}$ to account for systematically-lower values of
$\nu_{\mathrm low}$; note that
$f_{\mathrm lowE}$ is a strong function of the 
shape of the energy spectrum of relativistic electrons, and is
thus likely to vary systematically with $\alpha$. 
We use the symbol $u_{\mathrm base}$ to
denote the evaluation of equation~\ref{eq:miley}
with these assumed cut-off frequencies, with $\phi = 90^{\circ}$,
with $k = 0$, and with $\eta = 1$. This means that 
\begin{equation}
\label{eq:simplemiley}
u_{\mathrm min} =  \left[
(1+k) f_{\rm geom} f_{\rm lowE} / \eta \right]^{4/7} u_{\mathrm base}.
\end{equation}

The value of $u_{\mathrm base}$ can be calculated for a given source
(of known $z$) by estimating the surface brightness at 
the observed frequency $\nu$ from the total 
monochromatic flux density at that frequency divided by the 
solid angle subtended by the source, again
assuming a projected cylindrical geometry.

The time-averaged power of both jets $\bar{Q}$ is then given by
\begin{equation}
\label{eq:burbidge_over_falle}
\bar{Q}  =  g_{\mathrm exp} g_{\mathrm ke}  \left(
\frac{ \left[ (1+k) f_{\mathrm geom} f_{\mathrm lowE} 
/ \eta \right]^{4/7} f_{\mathrm min} u_{\mathrm base} \eta V}{t} \right)
\end{equation}
\noindent
where $f_{\mathrm min}$ is a factor which accounts for deviations
from the minimum energy condition
which corresponds to the energy density of the lobe magnetic field 
being about 75 per cent of the energy density of the relativistic particles.
In addition to internal energy density in the lobes, we have also
accounted for energy delivered by the jet, but converted into other forms:
$g_{\mathrm exp}$ is a factor which accounts for energy lost via the
expansion work done by the radio source; and 
$g_{\mathrm ke}$ accounts for the bulk and turbulent kinetic energy of 
the lobe plasma. As has been known since the work of
Longair, Ryle \& Scheuer (1973), 
it is safe to neglect the
gravitational potential energy 
of the lobe material. We have also neglected 
radiative losses, a point we will return to in 
Section 4.1.3. If we define
$U_{\mathrm base}$ as $u_{\mathrm base}$ times the
volume of the radio source, and absorb all the
unknown factors into $f$, we obtain
\begin{equation}
\label{eq:simple_burbidge_over_falle}
\bar{Q} =  \frac{f U_{\mathrm base}}{t}, 
\end{equation}

The most uncertain factor in equation~\ref{eq:burbidge_over_falle}
is the value of $k$. If we assume that the material in the 
lobes originated in the jet, that is we assume
that the radio lobes push back, and therefore exclude, any ambient material,
as has been argued fairly convincingly in the case of Cygnus A 
(Carilli, Perley \& Harris 1994), determining the value of $k$ 
is equivalent to determining the composition of the jet.
If the jet material and hence the 
synchrotron-emitting lobe plasma is entirely electrons and positrons then
$k=0$, whereas $k$ could, in principle, be very much higher
for an electron-proton plasma: the
simplest theories for particle acceleration via the
shock-Fermi process suggest $k \sim 10$ (Bell 1978), and higher
values are plausible (see Eilek \& Hughes 1991 for a review). 
For this reason, we consider two models: in model A we assume that 
the lobe material consists entirely of electrons and positrons;
in model B, protons make a significant, and perhaps dominant
contribution to the energy density of the lobes. Model A
is favoured by the recent detection of circular polarization in the 
radio cores of powerful radio sources (Wardle et al. 1998), as well as 
by an older dynamical argument concerning the 
requirements necessary for light jets to inflate lobes 
(e.g. Williams 1991).

It is possible to make an explicit estimate of the 
combined effects of the factors $g_{\mathrm exp}$
and $g_{\mathrm ke}$. If, as is normally assumed, 
the material that fills the lobes passed through the hotspot regions,
the points where the jets terminate, then it must 
have expanded from regions characterised by typical hotspot pressures 
$p_{\mathrm hs}$ to typical lobe pressures 
$p_{\mathrm lobe} \approx \frac{u_{\mathrm min}}{3}$.
For an adiabatic expansion, the value
of $(p_{\mathrm hs} / p_{\mathrm lobe})^x$, with $x \approx -1 / 4$,
measures the ratio of the internal energy of a 
volume element in the hotspot to that in the lobe
(equation 18 of BRW99).
If the ratio of these pressures scales with the 
radio surface brightness (see equation~\ref{eq:miley}) then
observations of typical hotspot and
lobe surface brightnesses in FRIIs, implies that of order 50 per cent of the
energy is retained as lobe internal energy, the remainder being converted 
to kinetic energy and/or lost as
work done by the expanding lobes; since the exponent $x$ 
is such a small fraction, this assumption is insensitive to
even fairly major departures from the minimum
energy condition in one or both of the 
hotspot/lobe regions (Leahy 1991 and 
Bicknell, Dopita \& O'Dea 1997 reach similar conclusions using 
similar arguments). We thus take
$g_{\mathrm exp} g_{\mathrm ke} \sim 2$\footnotemark.

\footnotetext{Note that larger expansion losses were
suggested by Kaiser, Dennett-Thorpe \& Alexander (1997) but that these
arose because of an error in interpretation giving incorrect estimates
of the energy lost via lobe expansion (Kaiser, priv. comm.).}

In the case of model A we can further argue that kinetic energy makes
such a small contribution to the energy budget that the internal
energy lost is converted almost entirely into work done on the ambient
material, that is $g_{\mathrm exp} \sim 2$ and $g_{\mathrm ke} \sim
1$.  Radio-emitting lobe plasma is known to be undergoing an ordered
backflow from the hotspot region (e.g.\ Miller 1985), and
spatially-resolved estimates of the ages of the lobe electrons suggest
backflow speeds of up to $V_{\mathrm bf} \sim 0.2 c$ (e.g. Liu, Pooley
\& Riley 1992).  The kinetic energy density of this bulk flow is
$\frac{1}{2} \rho V_{\mathrm bf}^2 \approx \frac{2}{9} u_{\mathrm min}
M_{\mathrm bf}^2$ (assuming a Mach number $M_{\mathrm bf}$, and a
relativistic equation of state for the lobe material). In model A the
sound speed will be $\sim c / \sqrt{3}$, so the backflow is at
most transonic ($M_{\mathrm bf} \sim 1$), and its kinetic energy
density a small, and for our purposes negligible, fraction of
$u_{\mathrm min}$.

Values significantly greater than unity for most of the other unknown
factors\footnotemark (most importantly $f_{\mathrm lowE}$, $1+k$ and 
$f_{\mathrm min}$) would lead to significant increases in 
internal energy density, and hence pressure, from the base 
value of $u_{\mathrm base}$. A general, if not water-tight, 
argument (e.g. Leahy 1991) suggests that one can put limits
on their combined effect by making a quantitative comparison 
of lobe pressures derived from application of
equation~\ref{eq:miley} 
with external pressures derived
by independent, typically X-ray based, techniques. 
In radio components bounded by weak subsonic shocks one expects that these
pressures should be roughly equal, so the observation (Feretti et al.
1995) that $u_{\mathrm base}/3$ can drop to $\sim 10$ per cent 
of the external pressure in the tails of ~head-tail' radio sources
suggests that 
\begin{equation}
\label{eq:pressureconstraint}
[(1+k) f_{\rm geom} f_{\rm lowE} / \eta]^{4/7} f_{\mathrm min} \ltsimeq 10.
\end{equation}
The limit in this equation follows from the
observation that, due perhaps to the
uncertainties introduced by entrainment, 
the minimum pressure in the relaxed radio structures of
FRI radio sources are often significantly closer to 
external X-ray derived pressures than those in the study of Feretti et al.
(e.g. Doe et al. 1995).
\footnotetext{
Note from eqn~\ref{eq:burbidge_over_falle} that $\eta$ is a parameter 
with a value less than unity, so that $u_{\mathrm base}$ 
will under-estimate the true internal energy density, 
but because the effective volume decreases linearly
with $\eta$, the net result is that $\bar{Q}$ will be over-estimated.
}
If one adopts a similar upward correction to the internal pressures of
the lobes of classical doubles, then they are typically at a much higher
pressure than their undisturbed environments: this is in 
keeping with the expectation that the lobes are
confined by an expanding sheath of shocked gas in ram-pressure
balance with the external medium (Scheuer 1974), and thus
in quantitative agreement with simple models for the lobes
of Cygnus A (Arnaud et al. 1984).

We can speculate further on some of the constituent parts of 
equation~\ref{eq:pressureconstraint}. 

\begin{itemize}

\item $f_{\mathrm geom}$. If
the magnetic field direction is randomly 
distributed with respect to the line of sight, 
$f_{\mathrm geom} \sim 1.4$. 

\item $f_{\mathrm lowE}$.
The circular polarization 
results of Wardle et al. (1998) suggest that the distribution of 
electron energies extends all the way down to mildly-relativistic
($\gamma \sim 1$) energies,
so that the imposition of a low-frequency cut-off at 10 MHz
(corresponding to $\gamma \sim 600$ for a lobe
magnetic field $B \sim 1~ \rm nT$)  
leads to a significant under-estimate of the energy density in
relativistic electrons. The value of
$f_{\mathrm lowE}$ depends critically on the slope $p$
of the electron energy distribution, and, recalling that $p = 2 \alpha + 1$,
on the radio spectral index: for
a $p$ of about 2, 
$f_{\mathrm lowE} \propto \ln (\nu_{\mathrm high} / \nu_{\mathrm low})$,
which extrapolating to the $\gamma \sim 1$ population 
implies $f_{\mathrm lowE} \sim 2$; for $p=3$, 
$f_{\mathrm lowE} \propto \nu_{\mathrm low}^{-0.5}$, so that 
$f_{\mathrm lowE} \sim 10$ becomes plausible. 
The former value is likely to be appropriate unless there
is evidence for a steep slope in the energy distribution 
from the low-frequency spectral index.
We will return to this
point in Section 4.3.

\item $f_{\mathrm min}$. 
Harris, Carilli \& Perley (1994) argue that the most likely explanation
of X-ray emission from the hotspots of Cygnus A is synchrotron
self-Compton (ssC) emission, which with $k=0$ (Model A) gives
magnetic field strengths very close to those estimated from the
minimum energy arguments. Details of the expansion
process from hotspots to lobes are required to
see if this equipartition of energy densities between the particles
and the magnetic field is preserved in the lobe. 
For a randomly oriented field and a relativistic equation of state the  
ratio will be preserved exactly, but if the magnetic field is 
ordered with respect to the flow lines then the diverging
flow from the hotspot, together with flux freezing, could cause the
energy density of the magnetic field to drop well below that of the 
particles, and $f_{\mathrm min} \sim 10$ becomes 
plausible (Alexander \& Pooley 1995), although it is
not clear whether such a situation is stable.
If $k \gg 0$ (Model B) then the correspondence between
the X-ray flux expected from the ssC process in an equipartition
magnetic field, and that observed from the hotspots of
Cygnus A must be coincidental, and $f_{\mathrm min}$ is unconstrained.

\item $\eta$. As emphasised by Leahy (1991), there are no hard
constraints on the volume filling factor.  Observations of filamentary
structure in the lobes of Cygnus A have been used to argue for $\eta
\sim 0.1$ (Perley, Dreher \& Cowan 1984; but see Tribble 1993).

\end{itemize}

Putting these considerations together for model A, 
equation~\ref{eq:pressureconstraint} can always be satisfied 
for $\eta \sim 1$ if the magnetic field and particle energy densities
diverge as a result of the expansion from hotspot to lobe, 
and/or if there is a large 
energy reservoir 
(such that $f_{\mathrm lowE} \sim 40$)
in mildly-relativistic electrons.
For $\eta \sim 0.05$ both equipartition and a low 
value for $f_{\mathrm lowE} \sim 2$ are perfectly acceptable.
In the former case, combination of all the relevant factors
in equation~\ref{eq:burbidge_over_falle} suggests $f \sim 20$, whereas
in the latter case $f \sim 1$. These probably bracket the 
uncertainties in model A.

For model B, values of $k \gta 20$ are ruled out by
equation~\ref{eq:pressureconstraint} (and would require a more
contrived explanation for the X-rays from the hotspots of Cygnus A) so
that possible combinations of $k$, $\eta$, $f_{\mathrm lowE}$ and
$f_{\rm min}$ again lead to the conclusion that $1 < f \lta 20$. In
this case the lower bound corresponds to $k \sim 1$, $\eta \sim 0.1$,
$f_{\mathrm lowE} \sim 2$ and $f_{\rm min} \sim 1$, and the upper
bound to $k \sim 20$, $\eta \sim 1$, $f_{\mathrm lowE} \sim 2$ and
$f_{\mathrm} \sim 1$. Note that the range of allowed values of $f$ is
the same in both models A and B: in either model large values of $f$
require well-filled lobes, and additionally in model A the lobe plasma
must either be far from equipartition or dominated by low-energy
electrons.  
Note that RS91 took $f \approx 2$ in equation
\ref{eq:simple_burbidge_over_falle}, which on the basis of the discussion of
this section, remains a reasonable estimate.

Finally, we briefly consider ways in which $\bar{Q}$ might
systematically under- or over-estimate $Q$, the instantaneous jet
power of the source. Assuming a time-independent
$Q$, BRW99 have presented a model for the 
development of an FRII radio source which suggests that
$p_{\rm hs} / p_{\rm lobe}$ increases with source age, and hence with $D$,
so that the ratio of work done to energy stored, and 
hence $Q$ to $\bar{Q}$ (as calculated with
a fixed $g_{exp} \sim 2$ in 
equation~\ref{eq:burbidge_over_falle}),
also increases systematically with $D$. However, the
low value of the exponent in equation 18 of BRW99 ensures that this
remains a relatively modest effect: adopting the BRW99 model
$\bar{Q} / Q$ might be expected to drop
by at most an order of magnitude over the range of 
ages, and hence $D$, spanned by FRII sources in complete samples. 
However, there is also a plausible mechanism for 
increasing $\bar{Q} / Q$ systematically with age.
In a simple model in which the ratio of work done
to energy stored remains constant but
$Q$ decreases monotonically with source age, $\bar{Q}$ 
would tend to over-estimate $Q$.

\subsubsection{Radio source environments}

Physical models for FRII radio sources predict that sources of
fixed $Q$ evolve differently with time in different gaseous environments 
(Scheuer 1974; Falle 1991). Any quantitative study of the 
FRII population must attempt to account for this effect. 
Unfortunately, the nature of the environments of 
FRII radio sources cannot be straightforwardly deduced from any
existing observational data. Even at very low redshifts ($z \sim 0.1$), 
X-ray observations currently lack both the sensitivity and 
resolution to map the thermal emission from
the gaseous haloes of FRII radio galaxies with the 
exception of a few, possibly atypical, sources like 
Cygnus A (Reynolds \& Fabian 1996). Other probes of the environmental 
density, including galaxy counting (e.g. Hill \& Lilly 1991),
weak gravitational lensing (e.g. Bower \& Smail 1997), and studies
of radio polarization (e.g. Garrington \& Conway 1991), are  
inherently indirect. We will discuss these various observational
constraints in detail elsewhere 
(Rawlings, Willott \& Blundell, in prep, hereafter RWB), 
and here merely state, and subsequently employ, various assumptions 
about typical FRII environments at both low and high redshift.

Following Falle (1991) and Kaiser \& Alexander (1997)
we will characterise the electron number density profile $n (r_{100})$
of the radio source environment by a power-law,
\begin{equation}
\label{eq:ka}
n (r_{100})=n_{100}  r_{100}^{-\beta},
\end{equation}
where $r_{100}$ is the radial distance from the active nucleus,
measured in units of $100 ~ \rm kpc$, and $n_{\mathrm 100}$ is the
electron number density when $r_{100} = 1$. The normalisation radius
of $100 ~ \rm kpc$ will be denoted as $a_{\rm o}$ in subsequent
calculations. This profile is a good
approximation only at radii $\sim 100 ~ \rm kpc$ since the
true gas density profile is likely to be
concave (e.g. Navarro, Frenk \& White 1997); a
detailed discussion, and arguments for a generic 
$\beta$ value of $1.5$, will be presented in RWB; 
here, we simply note that this profile is a good fit
to the $n(r)$ functions derived for the clusters embedding the
luminous radio sources Cygnus A and 3C 295 (Reynolds \& Fabian 1996;
Neumann 1999).

Estimation of the normalisation
parameter $n_{\mathrm 100}$ in equation~\ref{eq:ka} is more problematic.
At low redshift ($z \sim 0.1$) FRII radio galaxies are found to be 
associated almost exclusively with 
isolated elliptical galaxies and/or poor groups of galaxies, that
is environments less rich than 
Abell clusters (e.g. Prestage \& Peacock 1988, 1989). 
This suggests a typical value of $n_{100} \sim 400 ~ \rm m^{-3}$,
but to account for low-redshift environments as poor as an 
isolated elliptical, and as rich as 3C 295, it
can seemingly lie anywhere in the range $10^2 \ltsimeq n_{100}
\ltsimeq 10^4 ~ \rm m^{-3}$.

At higher redshifts ($z \sim 0.5-0.8$) galaxy-counting experiments
(Yates, Miller \& Peacock 1989; Hill \& Lilly 1991;
Allington-Smith et al. 1993; Roche et al 1999)
show that FRIIs, covering a wide range of
radio luminosity, are typically associated with rich groups, 
environments alternatively known as 
poor clusters.
The significant
fraction of X-ray detections amongst $z \sim 1$ 3C radio sources (Crawford
\& Fabian 1996), X-ray emission which in some cases is clearly
extended and probably thermal, suggests that this gradual increase in
environmental density persists to $z \sim 1$.  This trend with
redshift provides the most straightforward explanation for the
systematic increase in the depolarisation of the lobes of radio
quasars from redshift $z \sim 0$ 
to $z \sim 1$ (e.g. Garrington \& Conway 1991).
These results suggest a value of $n_{100} \sim 3000 ~ \rm m^{-3}$ 
as a typical normalisation value at $z \sim 1$.
Taken at face value the small scatter in the depolarisation data 
at high redshifts implies
a fairly small spread about this value of, say, $\pm 0.25$ dex, but 
since the published depolarisation work is largely confined
to the brightest (e.g. 3C) sources, this may be in part a 
selection effect. The analysis of Section 
4.1.3. is, however, confined to this same bright population, and so
$n_{100} = 3000 ~ \rm m^{-3}$ should be an appropriate 
normalisation factor; the consequences of systematic changes 
in environment with redshift will be discussed in 
Section 4.3.

\subsubsection{Ages and head advance speeds}

To determine jet powers and their relationship with the ages and sizes
of radio sources, we follow the dimensional analysis of Falle (1991).
He showed that the quantities $Q$, $\rho_{100}a_{\mathrm o}^ {\beta}$
(where $\rho_{100}$ is the mass density at 100 kpc equivalent to
electron density $n_{100}$), and {\em \.M$_{\mathrm J}$} (the jet mass
flux) can define a characteristic length scale, $L_{\mathrm o}$,
\begin{equation}
\label{eq:falledim}
L_{\mathrm o}=\left( \frac{ \left( \rho_{100}a_{\mathrm
o}^{\beta} \right)^{2} Q}{2 \dot{M}_{\mathrm J}^{3}}\right)^
\frac{1}{2\beta-4}.
\end{equation}
For typical FRII source parameters, $L_{\mathrm o}\sim 10$ pc, which
is small compared to the size of radio jets. At distances much greater
than $L_{\mathrm o}$, the mass in the bow shock is dominated by
material swept up by the jet, so {\em \.M$_{\mathrm J}$} is
unimportant. This leads to a bow-shock which
is self-similar, the future dynamics of which can be determined by a
simple dimensional analysis. Kaiser \& Alexander (1997) argued that
the self-similar growth also applies to the cocoon excavated by the
jet. BRW99 present a model in which the cocoon does
not grow self-similarly ($R_{T}$ and $g_{\mathrm exp}$ are slow
functions of source age). In all these models, the evolution of the
projected linear size of the radio source, $D$, is given by
\begin{equation}
\label{eq:falled}
\frac{D}{\sin\theta}=2c_{1}a_{\mathrm o}t^ {\frac{3}{5-\beta}} \left(
\frac{ Q} {2 a_{\mathrm o}^{5} \rho_{100}}\right)^
{\frac{1} {5-\beta}},
\end{equation}
where the factors of 2 come in because the linear size is the
separation between the ends of each jet of power $Q/2$. $c_{1}$ is
a constant unconstrained by the dimensional analysis. Re-arranging
this expression to make $t$ the subject tells us how the age of a
radio source is related to $D,Q$ and its environment;
\begin{equation}
\label{eq:fallet}
t = \left( \frac{D}{2\sin\theta c_{1}a_{\mathrm o}}\right) ^{\frac
{5-\beta}{3}} \left( \frac{2 a_{\mathrm o}^{5} \rho_{100}}
{Q} \right) ^{1/3}.
\end{equation}

To constrain the constant $c_{1}$ we now consider the instantaneous
head advance speeds of FRII radio sources. Using a method based on
the length asymmetry between the two lobes of radio sources, Scheuer
(1995) showed that radio source advance speeds derived by synchrotron
ageing methods (e.g. Alexander \& Leahy 1987, Liu et al.\
1992) are over-estimated due to large contributions to the
derived flow speeds by backflow. The samples of objects investigated 
by Scheuer are of the most powerful radio sources known: 
they have a mean radio luminosity of $\log_{10} 
(L_{151}$ / WHz$^{-1}$sr$^{-1}) =28.0$ and a median
projected linear size of 110 kpc. Since they are mostly quasars, we
take a typical angle to the line-of-sight of 45$^{\circ}$. This gives
a median de-projected linear size of 160 kpc -- the same as that of
the whole 3CRR and 7C sample. The mean head advance speed 
$V_{\mathrm head}$ derived
by Scheuer is $0.03c$, with a firm upper-bound of $0.15c$.
We adopt the typical `high-redshift' environment for an 
FRII discussed in Section 4.1.2.

Differentiating equation \ref{eq:falled} with respect to time
(keeping $Q$ constant), we obtain the advance speed of the radio
source, $v$,
\begin{equation}
\label{eq:fallev}
2v=\frac{dD}{dt}=2\sin\theta c_{1}a_{\mathrm o}\left( \frac{3}{5-
\beta} \right) t^{\left( \frac{\beta-2}{5-\beta}\right)} \left(
\frac{Q}{2 a_{\mathrm o}^{5} \rho_{100}} \right)
^{\frac{1} {5-\beta}}.
\end{equation}
Note the $t$-dependence of the advance speed. This shows that the
advance speed is constant throughout a source's lifetime only in a
$\beta=2$ environment. For $\beta=1.5$, we find that $v\propto
t^{-1/7}$, i.e. the source decelerates, albeit slowly, as it
grows. Substituting $D$ for $t$ from equation \ref{eq:fallet} gives us
an expression relating $v$ and $D$ for a particular value of $Q$. We
find $v\propto D^ {(\beta-2/3)}$ which is $v\propto D^ {-1/6}$ for
$\beta=1.5$.  Therefore, if we know the current hotspot advance speed
(inferred from Scheuer 1995), we can determine $v$ throughout the
source's lifetime and hence calculate the age of the source. We find
that for $v=0.03c$ and $v=0.15c$, the typical ages of these sources
are $7.4\times10^{6}$ and $1.5\times10^{6}$ years, respectively.
Equation \ref{eq:fallev} shows that, for a given $c_{1}$, $v$ and $D$,
there is a degeneracy between $Q$ and $t$. However, this degeneracy
can be broken because equation \ref{eq:simple_burbidge_over_falle} contains
another relationship between $Q$ and $t$.

\begin{figure*}
\epsfxsize=0.9\textwidth
%\begin{flushleft}
%\hspace{-1.3cm}
%\hspace{0.4cm}
\epsfbox{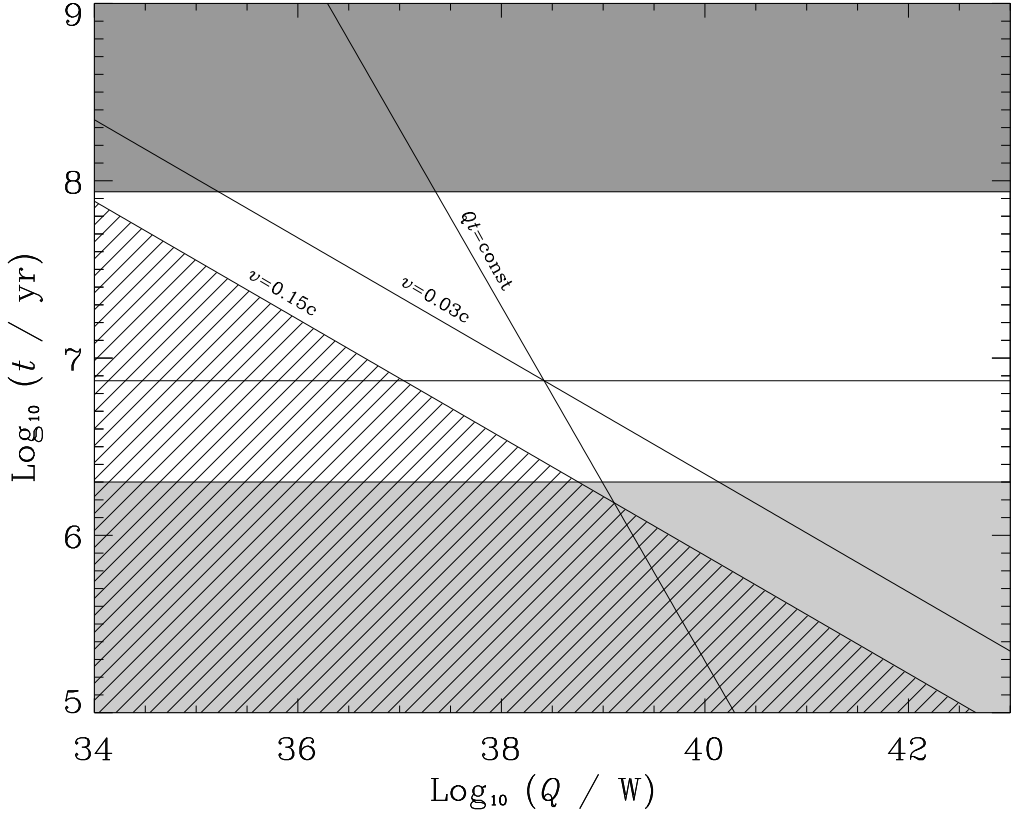}
%\end{flushleft}
{\caption[junk]{\label{fig:c1plot}$Q$ versus the age of the powerful
FRII sources investigated by Scheuer (1995). The line labelled $Q t$
is the energy conservation requirement of equation
\ref{eq:simple_burbidge_over_falle}. The lines labelled $v=0.03c$ and
$v=0.15c$ are from equation \ref{eq:fallev}, with values of $c_{1}$
such that they intersect the $Qt$ line at the ages derived in the
text. The pale shaded region shows the limit on $t$ set by the
observations of Liu et al.\ (1992). The dark shaded region shows the
limit on $t$ from the requirement that the jet be supersonic
throughout its lifetime.  The diagonal line shaded region is
disallowed by the observations of Scheuer (1995). The intersection of
the $Q t$ and $v=0.03c$ lines give our best estimate of the typical
ages and jet powers of the sources used by Scheuer; $t\approx 10^7$
years and $Q \approx 3\times10^{38}$ W. This diagram fixes the value
of the unknown constant $c_{1}$.}}
\end{figure*}
 
Inserting the ages calculated above into equation
\ref{eq:simple_burbidge_over_falle} gives a unique value of $Q$ for a source
of a particular age. Figure \ref{fig:c1plot} plots $Q$ against $t$
from equation \ref{eq:simple_burbidge_over_falle}. Equation
\ref{eq:simple_burbidge_over_falle} is simply an energy conservation line
from the minimum energy assumption. The two lines labelled $v=0.03c$
and $v=0.15c$ in Figure \ref{fig:c1plot} are from equation
\ref{eq:fallev}, given that they intersect the line from equation
\ref{eq:simple_burbidge_over_falle} at the ages calculated above.  They
therefore each give a value of the unknown constant $c_{1}$. The
region to the lower-left of the line labelled $v=0.15c$ is ruled out
by the upper limit to the hotspot advance speed (Scheuer 1995).

We also have two further constraints on the regions allowed in the
$Q$--$t$ plane. Spectral ages of powerful FRIIs have been determined
by Liu et al.\ (1992). The mean age for their most powerful sources
[$\log_{10} (L_{151}$ / WHz$^{-1}$sr$^{-1}) \geq 27.5$ and $D\approx
100$ kpc] is $2\times10^{6}$ yrs. Note that this is a lower limit to
their actual ages because of the arguments about the importance of
backflow put forward by Scheuer (1995). An upper limit to the ages of
FRII radio source comes from the requirement that the head of the
radio source advances supersonically throughout its lifetime. We
calculate the sound speed of the IGM assuming $T=2$ keV,
an appropriate value for a rich group. 
Assuming that the head advance speed equals the sound speed
throughout its lifetime, we derive an upper limit to the age of
$10^{8}$ yrs. Note that this is a strict upper limit because
we have already shown that the jet decelerates as it propagates. These
upper and lower age limits are shown in Fig. \ref{fig:c1plot}.

The intersection of the $Q t$ and $v=0.03c$ lines give our
best estimate of the typical ages and jet powers of the sources used
by Scheuer. Ages of $\approx 10^7$ years are consistent with our
strict upper and lower age limits derived above. Note that if the
current hotspot advance speeds are as high as $v=0.15c$, then the
calculated ages would be below the lower limit from spectral ageing.
The derived jet powers are $Q \approx 3\times10^{38}$ W.
From the intersection of the $Q t$ and $v=0.03c$ lines we
derive a value of the constant, $c_{1}$. 
This clearly depends critically on the value taken for $f$ 
in equation~\ref{eq:simple_burbidge_over_falle}: for $f=1$,
$c_{1}=5.4$ and for $f=20$, $c_{1}=2.3$.

Now that we have estimated $c_{1}$, it is trivial to estimate the jet
power. Substituting $t$ from equation \ref{eq:fallet} into equation
\ref{eq:simple_burbidge_over_falle} gives
\begin{equation}
\label{eq:falleq} 
{Q} = \left({f u_{\mathrm min}V} \right)^ {3/2}
\left( \frac{D}{2\sin\theta c_{1}a_{\mathrm o}}\right) ^{\frac
{\beta-5}{2}} \left( 2 a_{\mathrm o}^{5} \rho_{100} \right)^
{-1/2}.
\end{equation}
Since $u_{\mathrm min} \propto L_{151}^{4/7}$ from equation
\ref{eq:miley}, we find that $Q \propto L_{151}^{6/7}$ (at fixed
$D$). Hence this model predicts a relationship between the jet power
and the observed extended radio luminosity which is slightly less
steep than linear. $u_{\mathrm min}V \propto D^{9/7}$, which gives $Q
\propto D^{-4/7+\beta/2}$ (for sources with the same $L_{151}$)
\footnotemark.  Hence, for $\beta=1.5$, this gives a very weak
expected $D$ dependence of $Q \propto D^{\frac{5}{28}}$.  Note that
very small changes in $\beta$ determine whether the derived $Q$ goes
up or down with $D$ at a fixed $L_{151}$, but for realistic values of
$\beta$, $Q$ is not a strong function of $D$. Therefore we
find the relationship between $Q$ and $L_{151}$ in the adopted
environment is given by
\begin{equation}
\label{eq:ql} 
Q\simeq  3\times 10^{38} L_{151}^{6/7}~ {\rm W},
\end{equation}
where $L_{151}$ is in units of $10^{28}$ W Hz$^{-1}$ sr$^{-1}$.

\footnotetext{RS91 assumed that the sideways growth of the source was
determined by ram pressure balance such that $\rho_{100} v^{2}
\propto u_{\mathrm min}/V$. This leads to a $D$-dependence of the form
$Q \propto D^{-4/7}$. This is identical to the dependence
we find for a uniform external density ($\beta=0$) as assumed by
RS91. Note also that ram pressure balance models give the same radio
luminosity dependence of the jet power as we find, namely $Q
\propto L_{151}^{6/7}$. We find no systematic difference between
the estimates of $Q$ in this paper and values derived for
individual sources in RS91.}

We can now check one of our previous assumptions -- that radiation
losses are negligible compared to the time-averaged bulk kinetic power
of the jet. We calculate the total energy radiated by sources with
parameters similar to those investigated by Scheuer (1995). These
sources have $\log_{10} (L_{151}$ / WHz$^{-1}$sr$^{-1}) \approx 28.0$
and $D=110$ kpc. The evolutionary track of a source with these
parameters on the $L_{151}$-$D$ plane is approximated by a power-law,
$L_{151} \propto D^{\frac{-5}{24}}$, for $\beta=1.5$ and a constant
$Q$. To get a maximum value of the synchrotron energy radiated, we
assume a steep spectral index ($\alpha=1$), integrate 
the spectrum over a (rest-frame) frequency range 10 MHz to 100 GHz,
and then integrate over the source age.  Note that it
is possible that the synchrotron spectrum extends to lower frequencies
than those observable, but that for each decade in frequency that the
spectrum extends beyond this, the total energy radiated is increased
only by a quarter. We find that the total synchrotron energy radiated
over the sources lifetime is of the order of a few per cent of $Q
t$. We are therefore clearly justified in our assumption that radiation
losses make a negligible contribution to the total energy budget.  
This is not a surprising result because the 
synchrotron lifetimes of the low-energy electrons, where the
bulk of the lobe particle energy is stored, are much larger than the
ages over which a radio source with a typical 
evolutionary track keeps above the flux density limit of
a complete sample (BRW99).

\subsection{The physical meaning of narrow emission line luminosity}

\subsubsection{The excitation mechanism for the narrow emission lines}

Even though the extreme ultraviolet luminosities of certain
high-redshift radio galaxies are most plausibly explained by 
continuum emission from young hot stars, the high-excitation 
emission line spectra seen even in these cases demonstrate 
that the narrow emission lines are overwhelmingly excited by the 
central engine (e.g.\ Dey et al.\ 1997, and see McCarthy 1993 
for a general review). There are still, however, 
at least four distinct physical ways in which the central engine can
excite emission lines.

\begin{enumerate}

\item Quasar illumination. A broad (half-opening angle $\approx
50^{\circ}$) radiation cone from a central quasar ionises and heats
ambient material which cools via narrow-line radiation. In the
ionisation-bounded case, where individual optically-thick clouds
absorb and re-radiate all the incident flux, the narrow-line
luminosity is proportional to the photoionising quasar luminosity
$Q_{\rm phot}$ modulo a covering factor $CF$. The small spread in the
equivalent width of narrow emission lines in low-redshift quasars
(e.g. Miller et al. 1992) suggest a narrow range in $CF$ and no
systematic variations of $CF$ with $Q_{\rm phot}$.

\item Blazar illumination.  A narrow (half-opening angle $\approx
10^{\circ}$) beam of Doppler-boosted optical/UV synchrotron emission
ionises and heats material which cools via narrow-line emission. Such
beams pointed towards Earth explain most of the properties of the objects
known as blazars.

\item Jet-cloud collisions. Jets collide with ambient clouds
transferring some fraction of their mechanical power to thermal energy
via shocks driven into the clouds.  Subsequent cloud cooling produces
narrow line emission.

\item Bow-shock excitation. Expanding lobes of FRII radio sources do
work on the ambient material mediated by a strong bow shock.  
If any of the material passing through this shock has
sufficient time to cool to $\sim 10^{7} ~ \rm K$ then the bow shock
can become radiative, so that some fraction of the power delivered to
the bow-shock can emerge as bremsstrahlung and line emission. This
radiation excites a proportional amount of narrow emission
line luminosity provided the densest material in the bow shock has
cooled to $T \sim 10^{4-5} ~ \rm K$.

\end{enumerate}

We will hereafter neglect mechanisms (ii) and (iii) for exciting
narrow emission lines because the bulk of the narrow emission lines in
radio galaxies come typically either from nuclear regions, or from extended
regions which viewed from the active nucleus cover a solid angle far
larger than it is possible to achieve via mechanisms (ii) and (iii).
Nevertheless, there are individual radio galaxies in which one or both
of these `narrow-beam' mechanisms are likely to be in operation,
and which contribute significantly to the total narrow-line
emission: the extranuclear
emission line filaments in Centaurus A have been ascribed to both
blazar illumination (Morganti et al.\ 1991) and jet-cloud collisions
(Sutherland, Bicknell \& Dopita 1993); mechanisms (ii) and (iii) have
also been invoked for 3C 356 (Lacy \& Rawlings 1994) and PKS2152-699
(di Serego Alighieri et al. 1989).

Deciding between mechanisms (i) and (iv) is complicated by the
realisation that fast shocks produce a local photoionising source
which can mimic the ionisation/heating effects of a quasar power law
(e.g.\ Binette, Dopita \& Tuohy 1985), especially when only optical
line diagnostics are available (e.g.\ Allen, Dopita \& Tsvetanov
1998). Nevertheless, there is good observational evidence that
mechanism (i) is normally dominant: the major
arguments are as follows.

\begin{itemize}

\item
Accounting for differential reddening between the 
continuum/broad-line region and the narrow-line region,
powerful steep-spectrum radio-loud quasars have the same small spread in 
their (rest-frame) narrow-line equivalent widths 
as radio-quiet quasars (Miller et al.\ 1992). Amongst the 
steep-spectrum quasars there is no evidence for any dependence of equivalent 
width on quasar luminosity (Jackson \& Browne 1991), on redshift 
(at least out to $z \sim 2$; JR97), or on radio source
size (Hirst et al. 1999). It thus seems 
hard to escape the conclusion that the quasar continuum
luminosity is the principal factor which controls the narrow emission
line luminosity.

\item
Many powerful radio galaxies show polarimetric and spectropolarimetric
evidence for hidden quasar cones impinging on scattering media
coincident with the narrow-line emitting gas (e.g.\ Tran et al.\ 1998
and refs. therein).

\item 
In a study of the UV line ratios of a large sample of $z\sim2$ radio
galaxies Villar-Martin, Tadhunter \& Clark (1997) concluded that the
ionisation states are primarily determined by photoionisation from a
central source.  Detailed models of photoionising shocks (e.g.\ Allen
et al.\ 1998) suggest that the UV collisionally-excited lines should
be strongly boosted in the high-temperature cooling zones behind fast
shocks, whereas the line ratios in $z \sim 2$ radio galaxies show no
evidence of this boosting and follow closely those predicted by a
quasar-illumination model.

\item
For objects confined to a narrow region of the
$L_{151}$ versus $z$ plane, the sizeable spread in 
narrow line luminosity seems well correlated with the
underlying quasar luminosity. The recent work of 
Simpson et al.\ (1999) illustrates this effect: 
they found an intrinsically ultraluminous thermal-infrared
nucleus in 3C 265, a $z \sim 1$ 3C radio galaxy with 
ultraluminous narrow emission lines, but
evidence for a weak thermal-infrared  nucleus in 3C 65, 
the $z \sim 1$ 3C radio galaxy with the
weakest narrow emission lines. The 
low-excitation line ratios of the weakest narrow-line emitters
provides indirect evidence that they harbour
intrinsically weak quasar nuclei (e.g.\ Tadhunter et al.\
1998). Evidence of line excitation correlating positively with
narrow-line luminosity was used by Saunders et al.\ (1989) to argue
that the narrow-line luminosity is proportional to the underlying
quasar luminosity in low-redshift FRII radio sources.

\end{itemize}

Evidence for the dominance of mechanism (iv) is confined to
individual, possibly atypical, cases, and is largely 
circumstantial. It is clear, however, that shocks in general, and the
bow-shock in particular, have a huge influence on the {\it
spatial structure} and {\it kinematics} of the extended narrow-line
emitting regions (e.g.\ McCarthy 1993). The key point here is that
even in a pure quasar-illumination picture, the presence of shocks can
have an important influence on the emergent line ratios, and some
impact on the line luminosities.  As the excitation level is
controlled by the effective ionisation parameter $U_{e}$ (the ratio of
the photon density to the electron density), and as densities are
typically boosted by shock compression, the ratio of high- to
low-excitation lines (e.g.\ [OIII] to [OII]) will, for a fixed
incident photon flux, be lower in post-shock gas (Clark et al. 1997;
Lacy et al. 1998). Cooling behind shocks may also introduce cool
UV-opaque material at large radii which raises the $CF$ and hence the
line luminosity.  For these reasons, some authors (e.g. Tadhunter et
al.\ 1998) invoke shocks to explain why some powerful narrow-line
radio galaxies have ultraluminous emission lines, but relatively low
excitation spectra (JR97). 
We note that it might be possible 
to obtain a more accurate indicator of the underlying 
quasar luminosity by restricting attention to emission line 
luminosity arising from the nuclear regions.
We will return to this important question
in Section 4.3, assuming for now that mechanism (i) is the
dominant way in which the AGN delivers energy to the narrow line region.

\subsubsection{The relationship between $L_{\mathrm [OII]}$ and 
$Q_{\rm phot}$}
 
Miller et al. (1992) showed that in a complete sample of the brightest
low-redshift ($z < 0.5$) optically-selected quasars, the narrow
emission line equivalent widths show very little scatter (0.2 dex) and
are not correlated with optical luminosity. This quasar sample was
taken from the Bright Quasar Survey (BQS; Schmidt \& Green 1983) which
because of its UV-excess selection criterion is dominated by `naked'
quasars in which differential reddening between the lines-of-sight to
the narrow-line region and the continuum producing regions should be
unimportant.  Within the Miller et al. BQS sub-sample there is no
evidence of any difference between the [OII] EW distributions of
radio-loud and radio-quiet quasars.  Differential reddening is a
likely explanation for the larger means and dispersions for the
distributions in [OII] equivalent widths in radio-selected samples of
quasars (Jackson \& Browne 1991; Baker 1997).  We assume that (before
any reddening) all steep-spectrum radio sources have rest-frame [OII]
equivalent widths of 10 \AA, which is the mean value determined by
Miller et al.

Given an [OII] line luminosity and an assumed EW we can calculate the
monochromatic blue luminosity.  To convert this to a bolometric
luminosity we use the Sanders et al. (1989) study of the spectral
energy distributions of BQS quasars. Sanders et al. observe a small
range in $L_{\mathrm bol}/ \nu L_{\nu} (B)$ in their sample, with a
mean of 16.5. We take $Q_{\mathrm phot}=L_{\mathrm bol}$ because the
big blue bump contributes at least 70\% of the bolometric luminosity
of quasars and at least some of the remainder (chiefly in the IR bump)
may be reprocessed emission from the accreting material. With an
assumed rest-frame [OII] equivalent width of 10 \AA, this gives
\begin{equation}
Q_{\mathrm phot} \simeq 5\times 10^{3} L_{\mathrm [OII]}~{\rm W}.
\end{equation}
RS91 estimated $Q_{\mathrm phot}$ by converting from one or two
observed narrow line luminosities to $L_{\rm NLR}$ and then assumed a
covering factor of $CF\sim 0.01$. Using the line ratios in McCarthy
(1993) we estimate that $L_{\rm NLR} \approx 15 L_{\mathrm [OII]}$ and
the inferred covering factors ($CF\sim0.003$) are similar to those of
RS91. 

\subsection{The relationship between $Q$ and $Q_{\rm phot}$}

In Section 4.1 we argued that the range in extended radio luminosity
spanned by FRII radio sources in complete samples is controlled
principally by a range in $Q$, the jet power, and gave a formula
mapping $L_{151}$ to $Q$. In Section 4.2 we argued that the range in
narrow emission line luminosity of FRII radio sources is controlled
principally by a range in $Q_{\rm phot}$, and gave a formula mapping
$L_{\rm [OII]}$ to $Q_{\rm phot}$.  In this section we put these results
together to investigate the origin of the emission line -- radio
luminosity correlation.

\begin{figure*}
\epsfxsize=0.9\textwidth
%\begin{flushleft}
%\hspace{-1.3cm}
%\hspace{-0.5cm}
\epsfbox{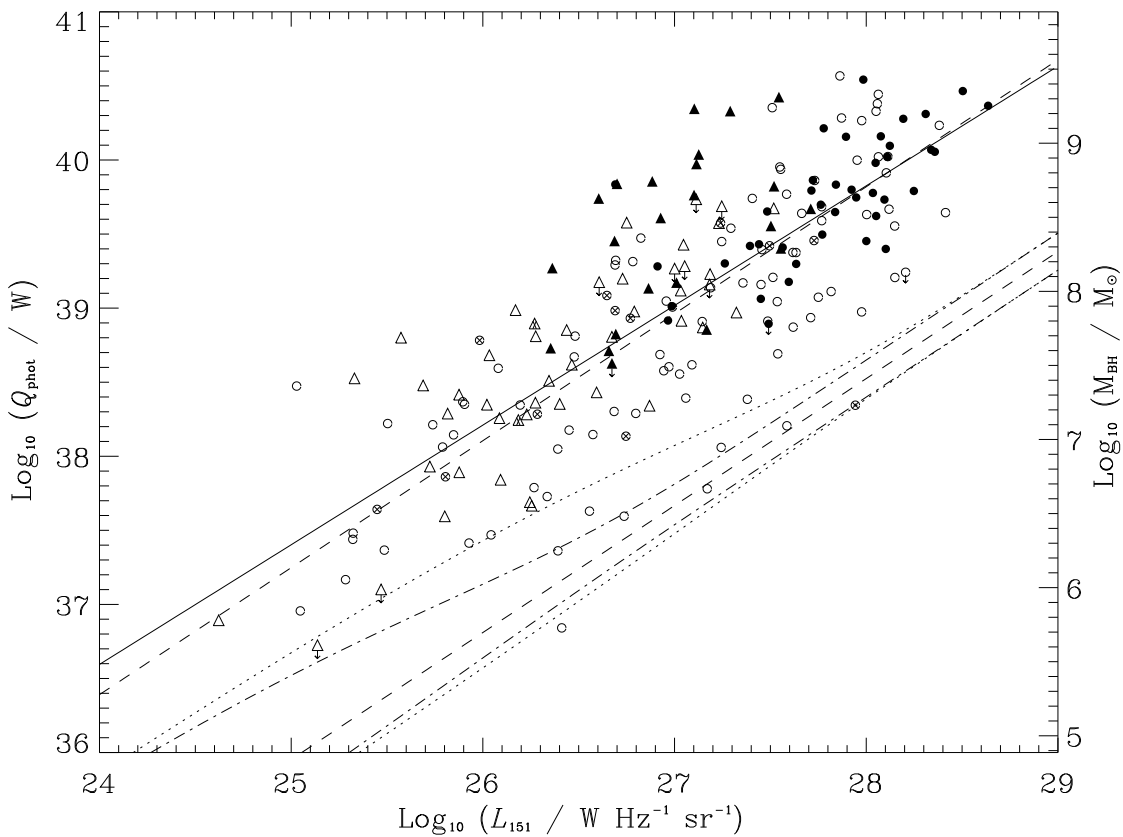}
%\end{flushleft}
{\caption[junk]{\label{fig:model}The $L_{\mathrm [OII]}$ versus
$L_{151}$ correlation of Fig. \ref{fig:loiilr}, where the vertical
axis now shows the photoionising power, $Q_{\mathrm phot}$, instead of
$L_{\mathrm [OII]}$. $Q_{\mathrm phot}$ is calculated for each source
as described in Section 4.2. The solid line is the least-squares fit
to all FRIIs of Section 3. The dashed lines show the variation of
$Q_{\mathrm phot}$ with $L_{151}$ as calculated in Section 4 assuming
$Q_{\mathrm phot}=Q$. The upper dashed line is for the case of $f=20$
and the lower one for $f=1$. The dotted and dot-dashed lines show the
uncertainty in the $f=1$ case due to a range of environments as
discussed in the text. The right vertical axis shows the central
black hole masses implied by this model for Eddington-limited
accretion (which we argue in Section 4.3 is unlikely to be the case
for objects with $Q_{\mathrm phot} \ltsimeq 10^{39} {\mathrm W}$).}}
\end{figure*}

We begin with the hypothesis that $Q_{\rm phot}$ is equal to the jet
power $Q$. This form of behaviour is characteristic of some symbiotic
jet-disc models (e.g. Falcke \& Biermann 1995). Figure \ref{fig:model}
shows the emission line--radio correlation (Fig. \ref{fig:loiilr})
again, with $L_{\mathrm [OII]}$ replaced by $Q_{\mathrm phot}$ (the solid
line calculated according to the
prescription of Section 4.2.2). 
Also on this figure are lines representing the relationship
between $Q$ and $L_{151}$ derived in section 4.1 (equation
\ref{eq:ql}) under the assumption that $Q_{\rm phot}=Q$. The first
thing to notice is that the gradients of the model lines (dashed) are
very similar to the gradient of the best-fit line. The model predicts
$Q \propto L_{151}^{0.86}$ and the fitted line gives $Q \propto
L_{151}^{0.81\pm0.04}$. The upper dashed line is for the case of
$f=20$ (which recall from section 4.1.1 is the maximum possible energy
according to the constraint of equation \ref{eq:pressureconstraint})
and the lower dashed line for the absolute minimum energy case of
$f=1$. We see that for $f=20$ then $Q_{\rm phot}$ and $Q$ are indeed
approximately equal, whereas for $f=1$, $Q_{\rm phot} \approx 20Q$,
i.e. the jets only carry 5\% as much power as is released during
accretion. If one instead assumes an open universe with $q_{0}=0$, then
$Q_{\mathrm phot} \approx 2 Q$ for $f=20$ and 40$Q$ for $f=1$. Hence
the ratio of jet to accretion power is half that for $q_{0}=0.5$. The
slope of the best-fitting $L_{\mathrm [OII]}$--$L_{151}$ correlation
(Section 3.1) for $q_{0}=0$ is $0.85 \pm 0.04$, i.e. still consistent
with the model slope of 0.86. 

Dunlop \& Peacock (1993) raised the possibility that the correlation
between the radio luminosity, $L_{\mathrm rad}$, and the narrow
emission line luminosity, $L_{\mathrm NLR}$, is actually due to
environmental effects. They noted, as we can see from equation
\ref{eq:falleq}, that for a given jet power, sources in higher density
environments have higher radio luminosities ($L_{151} \propto
\rho_{100}^ {1/2}$). Given the wide variety of environments in which
FRII sources are found, this is indeed likely to be a major cause of
scatter in the emission line -- radio correlation, but could it
dominate the correlation? 

To investigate environmental effects on the slope, normalisation and
scatter of the correlation, we adopt the range of environments at both
low- and high-redshift discussed in section 4.1.2. 
Recall that there is some evidence that the
scatter in environments 
at low- to moderate-redshift (from isolated ellipticals 
to clusters like that associated with 3C 295) is
greater than that at $z \gtsimeq 1$. Due to the tight correlation
between $L_{151}$ and $z$ in flux-limited samples, a mapping between
$L_{151}$ and a typical $z$ was possible for each of the 3CRR and 7C
samples. We then recalculated the $f=1$ model line assuming evolution
in the upper and lower limiting density environments.
Fig. \ref{fig:model} plots these bounds for the 3CRR (dotted) and 7C
(dot-dashed) samples. It can be seen that at high luminosities the
range of scatter due to environments is small, but at lower
luminosities it could be responsible for about one dex of scatter. 
The strongest systematic effect of environment we can envisage would
arise if all low power FRIIs were in isolated ellipticals and all high
power FRIIs were in rich clusters. This systematic change in environment
would flatten the expected $L_{\mathrm [OII]} - L_{151}$ correlation
to a slope of $\approx 0.6$. 

The emission line -- radio correlation holds over four orders of
magnitude, so we can rule out environmental effects as the cause of
the correlation, since they can account for only a maximum difference
in $L_{151}$ of one dex, 
corresponding to two dex in $n_{100}$
at a given $Q$. Moreover, Hill
\& Lilly (1991) do not find evidence for any correlation of $L_{\mathrm
rad}$ with clustering environment in their study of radio galaxies covering
a wide range of radio luminosities at $z\sim0.5$.  
Large-scale environmental effects cannot, of course, explain
correlations between the extended-radio and optical luminosities in
quasars (Serjeant et al. 1998; Willott et al. 1998a) since the optical
luminosity comes from nuclear regions. We therefore conclude that
environmental effects make a significant contribution only to the {\em
scatter} in the correlation. 

In equation \ref{eq:falleq} we showed that values of jet power derived
from the minimum energy condition and our model of source evolution
depends only weakly on the projected linear size $D$. Indeed,
Fig. \ref{fig:loiid} shows that there is little correlation between
the residual of the $L_{\mathrm [OII]}$ vs $L_{151}$ relation and $D$,
consistent with this model. In the model we find the expected $D$
dependence of $Q$ is a power-law with index ($-4/7+\beta /2$), which
for $\beta=1.5$ equals 5/28. The data shows a weak anti-correlation
between $L_{\mathrm NLR}$ and $D$. As mentioned in section 4.1.1, it
is uncertain whether $Q/ \bar Q$ increases with $D$ due to an increase
in the ratio of work done to energy stored, or whether $Q/ \bar Q$
decreases with $D$ due to $Q$ dropping with time, due perhaps to a
diminishing fuel supply. The finding in
section 3.2 that virtually all the relatively weak lined objects
[residual $\Delta \log_{10} (L_{\mathrm [OII]})<-0.5$] have $D \gta 100$ kpc,
may suggest the latter effect dominates, but it is impossible for us
to draw any firm conclusions here. One major reason for this is that 
explanations of such subtle effects cannot ignore the
possible influence of shocks on line luminosities as discussed in
Section 4.2.1. Indeed, the weak anti-correlation between (residual)
$L_{\mathrm [OII]}$ and $D$ may be the product of just such effects:
in sources of small $D$ the bow-shocks are within the
extended narrow-line region, and emission line luminosities
might be raised by compression, covering factor, and possibly
shock ionisation effects.

Fig. \ref{fig:loiialp} showed that there was no correlation between
low-frequency spectral index and (residual) emission line luminosity in the
complete samples. If a large fraction of the jet energy was in
low-energy electrons (i.e. $f_{\mathrm lowE} \gtsimeq 10$), then one
would expect objects with steeper low-frequency spectral indices
to have significantly higher total jet powers than those with flatter
spectra. Hence one would expect to see a positive correlation between
narrow line emission and spectral index. The lack of such a correlation
is thus consistent with a small value for $f_{\mathrm lowE}$.

The greatest effect which dominates the normalisation of the model is
the value of the factor $f$ which encompasses amongst other things the
nature of the composition of the jet, the filling factor, and the
low-energy cut-off. As discussed in section 4.1.1, it is most likely
that $f$ lies somewhere in the range between 1 and 20, the
consequences of which are plotted on Fig. \ref{fig:model}. Note that
$f$ does not affect the slope of the model, since there are likely to
be few systematic variations of these factors with jet power. Also, it
should contribute little to the scatter in Fig. \ref{fig:model}, since
the value of $f$ should not vary too widely from source to source
(i.e. jet composition is unlikely to be different in different
sources).

We re-emphasise the specific problem of the shock-related boosting of
narrow-line luminosity (for a given underlying quasar luminosity) 
discussed in Sec 4.2.1: correcting for such effects would shift
points vertically downwards in Fig. 7, but since the boosting factor
may be a function of $Q$, this might recover an underlying
correlation which has a significantly different slope to that 
evident in the raw data (e.g. Fig. 1). There may also be some effect on the
overall normalisation, and hence a small systematic increase in 
the inferred ratio of $Q$ to $Q_{\rm phot}$. Large systematic 
increases, that is order-of-magnitude effects 
similar to those caused by our ignorance of the jet composition (i.e. the
value of $f$), would require that shock-induced photoionisation
is the dominant source of narrow-line excitation. 
This would lead to the problems discussed in Sec 4.2.1
as well as an additional energetic difficulty made apparent by the
discussion of this section:  essentially all of the
power delivered to the bow shock (at most $\sim Q / 2$, see Sec 4.1.1)
would need to be converted into emission line radiation, requiring
highly efficient conversion of work done by the radio source
into photoionising radiation, and effective covering factors of order unity.

Factors contributing to the scatter about the mean radio--emission
line correlation will include: (a) intrinsic scatter in the underlying
$Q$, $Q_{\mathrm phot}$ relation to be discussed in Section 4.4; (b)
source-to-source variations in environment, value of $R_T$ (see
Leahy, Muxlow \& Stephens, 1989) as well as the various components of
the $f$ factor discussed in Section 4.1.1; (c) long-term time
variability effects, since the light-travel time to the narrow-line
region is much less than to the radio lobes; (d) emission lines
excited by means other than the radiation cone from the quasar,
e.g. shocks, blazar beams and possibly starbursts; and (e) use of
emission line ratios, and for some of the quasars, line equivalent widths, to
estimate [OII] strengths from other data.  Note that the weak residual
correlation between $L_{\mathrm [OII]}$ and $z$ may be the result of
some of these effects.

\subsection{Interpretation of the accretion--jet power link}

Our interpretation of Fig. \ref{fig:model} will follow many of the
arguments given previously by RS91. Given the strong correlation
evident in Fig. \ref{fig:model}, we will use the terms high- or
low-$Q$ to refer to FRII sources near the top-right and bottom-left of
the plot, respectively.  In the discussion below we will assume that
\begin{enumerate}
\item $Q$, the jet power, can be inferred from
$L_{151}$, as described in Section 4.1;
\item $Q_{\mathrm phot}$ (derived from $L_{\mathrm [OII]}$ as in
Section 4.2) informs directly on the accretion rate, $\dot{M}$, via
$\dot{M}= Q_{\mathrm phot}/ (\epsilon_{\mathrm phot} c^{2})$, such that
$\epsilon_{\mathrm phot}$ is the radiative efficiency of the accretion
process;
\item the Eddington luminosity of a black hole of mass $M_{\mathrm
BH}$ [$L_{\mathrm Edd}=1.3\times10^{31} (M_{\mathrm BH}/M_{\odot})$ W]
provides a natural upper limit to $\dot{M}$ via $\dot{M}\ltsimeq
L_{\mathrm Edd} /(\epsilon_{\mathrm phot} c^{2})$, and that for
Eddington-limited accretion $Q_{\mathrm phot}\simeq L_{\mathrm Edd}$;
\item for Eddington-limited accretion, the radiative efficiency,
$\epsilon_{\mathrm phot}$, corresponds to a unique timescale, $\tau
\approx 4.4 \times 10^{8}\epsilon_{\mathrm phot}$ yr. $\tau$ is simply
the mass of the black hole divided by its accretion rate, i.e. the
time required to accrete a mass equivalent to $M_{\mathrm
BH}$. Therefore if the source accretes at its Eddington luminosity for
a time comparable to $\tau$, its mass will increase significantly
during its lifetime. For an efficiency of 0.05 (similar to the
theoretical prediction of Shapiro \& Teukolsky 1983 for a non-rotating
black hole), this gives a timescale of $2\times 10^{7}$ yr.
\end{enumerate}

The observed link between accretion rate and jet power means that the
two processes are related in some way. The possibilities fall into two
broad categories: (i) if all FRII radio sources are accreting at
approximately their Eddington rates, then the central black hole mass
scales with the jet power; (ii) alternatively, there may be only a
small range in $M_{\mathrm BH}$ for FRII radio sources 
so that the low-$Q$ sources, at least,
may be accreting at sub-Eddington rates. We now consider the evidence
for each of these two possibilities.

Accounting for Doppler boosting and gravitational lensing in the most
extreme objects, the most optically luminous (and optically-selected)
quasars known have $Q_{\mathrm phot} \ltsimeq 10^{41}$ W (e.g.
Kennefick, Djorgovski \& De Carvalho 1995). Hence they overlap with
the high-$Q$ end of the sources plotted in Fig. \ref{fig:model}. One
would expect that this upper limit to quasar luminosities corresponds
to quasars which are accreting at their Eddington limits. The inferred
black hole masses for these Eddington luminosities are $\approx 10^{9-10}
M_{\odot}$. Note that if these objects are sub-Eddington accretors,
then their black hole masses would have to be such that $M_{\mathrm
BH} \gg 10^{9} M_{\odot}$. 

We conclude that the most powerful (high-$Q$) radio sources are
associated with black holes with $M_{\mathrm BH} \gtsimeq 10^{9}
M_{\odot}$, accreting at or near their Eddington limits ($\dot{M}
\approx 10 /\epsilon_{\mathrm phot} ~M_{\odot} ~{\mathrm yr}^{-1}$).

If all FRII radio sources are accreting at rates approaching their
Eddington limits, the power output of a radio-loud AGN (both optically
and in power in the jets) is a measure of its black hole mass. Hence
the ranges in $Q_{\mathrm phot}$ and $Q$ observed
corresponds to a range of $M_{\mathrm BH}$. On the right-hand vertical
axis of Fig. \ref{fig:model}, we plot the mass of a black hole
accreting at the Eddington luminosity as a function of $Q_{\mathrm
phot}$. The range of black hole masses for FRII radio sources
predicted with this model is $M_{\mathrm BH} \sim 10^{5.5}-10^{9.5}
M_{\odot}$. Note that nearly all the sources classified as quasars
have implied black hole masses $M_{\mathrm BH} \gtsimeq 10^{7.5} M_{\odot}$.

This range of $M_{\mathrm BH}$ is in good agreement with the masses of
possible remnant black holes in galaxies in the local Universe. For
example, studies of the gas dynamics of the inner regions of M87 imply
a supermassive black hole with $M_{\mathrm BH}= 3\times 10^{9}
M_{\odot}$, (Ford et al. 1996; Maccheto et al. 1997). It is
interesting to note that the highest power radio sources, which (with
the exception of Cygnus A) are all at high redshifts ($z>1$), appear
to have similar black hole masses to the black holes in local
brightest cluster galaxies. At the other end of the range, the
quiescent nearby galaxy M32 has recently been shown to harbour a dense
mass concentration of $3 \times 10^{6} M_{\odot}$ at its nucleus, most
probably as a black hole (van der Marel et al. 1997). Kormendy \&
Richstone (1995) found that the mass of these remnant black holes in
nearby galaxies are roughly proportional to the masses of their
spheroidal components. It should be noted, therefore, that all the
elliptical galaxies with absolute magnitudes brighter than $M_{B}=-19$
(which is the lower limit to FRII radio source host galaxy
luminosities; Owen \& Laing 1989) have remnant black holes in the
range $M_{\mathrm BH}= 10^{7.5}-10^{9.5} M_{\odot}$
(Franceschini, Vercellone \& Fabian 1998).

RS91 pointed out that the energy stored in the lobes of some radio
sources is so great that their black holes must have masses
$M_{\mathrm BH} \gg 10^{7} M_{\odot}$. This is particularly a problem
for large (and therefore old) FRII sources. The largest sources in our
sample have $L_{151}\approx 10^{26}$ W Hz$^{-1}$ sr$^{-1}$, ages 
$\gtsimeq 10^{8}$ yr and stored energies of $\sim 10^{53}$ J. Fig.
\ref{fig:model} shows that these sources typically have $Q_{\mathrm
phot} \sim 10^{38}$ W. Therefore if they are accreting at the
Eddington limit, their black hole masses would be only $M_{\mathrm BH}
\sim 10^{7} M_{\odot}$. For an accretion efficiency of
$\epsilon_{\mathrm phot}=0.05$, the total mass accreted by the black
hole during the sources lifetime would be equivalent to the amount of
energy stored in the lobes. This would seem to be an implausibly
efficient mechanism of feeding the lobes.

This problem is related to the accretion timescale problem mentioned
at the beginning of this section.  Some low-$Q$ sources have ages much
larger than the characteristic accretion timescale, so the effect of
this should be observable as a positive correlation between the age
(or, approximately, the size $D$) and the optical/radio luminosity of
low-$Q$ FRII sources. No such correlation has been observed (Section
3.2). Note that this problem can be avoided for most radio sources if
one assumes a greater efficiency (such as 0.4, which is possible for a
maximally rotating black hole; Shapiro \& Teukolsky 1983), in which
case $\tau=2\times 10^{8}$ yr. However, considering also the
theoretical (Efstathiou \& Rees 1988) and observational (Franceschini
et al. 1998) evidence for a correlation between host galaxy luminosity
and black hole mass, it is most plausible that low-$Q$ FRII sources
have black hole masses of at least $\sim 10^{8} M_{\odot}$ and are
accreting at sub-Eddington rates. Note that the high-$Q$ sources which
make it into a given radio survey are generally much younger than
$\tau$ (see BRW99), and therefore can easily be Eddington accretors
throughout their observable lifetimes, and thus avoid the timescale
and stored energy problems.

The evidence suggests that low-$Q$ sources have black hole masses
of $\gtsimeq 10^{8} M_{\odot}$ and are accreting at sub-Eddington rates.

We now consider whether there is any observational evidence for a
significant
range of black hole masses in quasars. Joly (1987) and Miller et
al. (1992) observed a positive correlation between the widths of
optically-selected quasar broad emission lines and their optical
luminosities, $L_{\mathrm opt}$. This correlation is consistent with
$M_{\mathrm BH} \propto L_{\mathrm opt}$, assuming the broad-line
emitting clouds are gravitationally bound, and moving on Keplerian
orbits. Koratkar \& Gaskell (1991) studied the broad emission line
variability of a sample of low-luminosity quasars and Seyfert Is to
infer the central black hole masses and radii of the broad line
regions. The black hole masses derived are $10^{8}-10^{9} M_{\odot}$
with the objects accreting at approximately 5\% of their Eddington
luminosities, further evidence that low-$Q_{\mathrm phot}$ objects are
not Eddington accretors with $10^{6}-10^{7} M_{\odot}$ black
holes. They observed a positive correlation between the bolometric
luminosity, $L_{\mathrm bol}$, and black hole mass of the form
$L_{\mathrm bol} \propto M_{\mathrm BH}^{1.1 \pm 0.3}$. 
Note, however, that the quasar samples studied to date by such
methods are dominated by radio-quiet objects, and although
the few radio-loud objects included obey the same general trends
there remains the possibility that studies of large samples
of radio-loud quasars will reveal a different behaviour.

We conclude that the black hole mass may scale with $Q_{\mathrm
phot}$ over a restricted part of Fig. \ref{fig:model}, but the main
difference between high-$Q$ and low-$Q$ sources is that the high-$Q$
sources are accreting near their Eddington luminosities, whereas the
low-$Q$ sources are sub-Eddington accretors.

Fig. \ref{fig:model} shows that if there is a range of black hole
masses then high-$Q$ sources, operating in the Eddington-limited
regime, should exhibit a scaling between radio luminosity and
$M_{\mathrm BH}$. If the central black hole mass of radio sources
scales with host galaxy mass (as is observed for nearby quiescent
galaxies, Kormendy \& Richstone 1995; and expected in some theories of
galaxy formation, e.g. Efstathiou \& Rees 1988, Silk \& Rees 1998),
then one would expect there to be a positive correlation between radio
luminosity and mass of the host galaxy.
Roche, Eales \& Rawlings (1998) find that 6C radio galaxies at
$z\sim 1$ are smaller and fainter than 3CR galaxies (Best, Longair \&
R\"ottgering 1998) at the same redshift (which are a factor of about 6
times greater in radio luminosity). They conclude that there is indeed a
positive correlation between the radio luminosity and host galaxy
luminosity. The lack of such correlations at low redshift 
is unsurprising given that, with few exceptions like 
Cygnus A, the FRIIs are low-$Q$ objects operating in
the sub-Eddington regime. 

At low redshifts, FRI and FRII radio galaxies are predominantly found
in different environments, the FRIs in clusters and FRIIs in isolated
ellipticals or poor groups (e.g. Prestage \& Peacock 1988). At higher
redshifts few FRI sources make it into
complete samples, and their environments are
largely unstudied, whereas, as discussed in 
Section 4.1.3, FRIIs typically apparently inhabit 
rich groups. Hence, it is possible that an ultramassive black
hole that, due to a high accretion rate, would produce a powerful FRII
radio source at high redshift would, because of a relatively meager
fuel supply, develop a low power FRI radio source at low redshift.  A
global decrease in available fuel for accretion is expected from
$z\sim2$ to the present day, due to the virialisation of galaxies and
clusters (e.g. Rees 1990). 
For example M87 now hosts an FRI radio source with $L_{151} \approx
10^{25}$ W Hz$^{-1}$ sr$^{-1}$ and $Q_{\mathrm phot}\approx 10^{37}$
W, thus putting it at the lower-left corner of Fig.
\ref{fig:model}. However, the dynamical evidence is that it contains
one of the most massive black holes known, $M_{\mathrm BH}= 3\times
10^{9} M_{\odot}$. Therefore it must be accreting at well below its
Eddington limit. 

Fig. \ref{fig:model} shows a close relationship between the jet power
and accretion rate, with $Q / Q_{\mathrm phot} \gtsimeq0.1$ over the
full range of $Q$. This is evidence for a single mechanism linking
these two properties over a large range of accretion rates
(e.g. RS91). Celotti, Padovani \& Ghisellini (1997) found a similar
relationship of $Q \approx Q_{\mathrm phot}$ by considering the broad
emission line luminosities and the power in parsec-scale jets of
quasars. They suggest that it is the magnetic field near to the black
hole which controls both the accretion luminosity and the jet power.
A symbiotic link between accretion and jets in radio-loud AGN such as
this has been proposed by Falcke \& Biermann (1995). In their model,
accretion energy is primarily dissipated by heating in the outer part
of the disc ($Q_{\mathrm phot}$), but primarily by the jet in the
inner part of the disc ($Q$). This model requires a substantial amount
of energy existing in the form of magnetic fields and relativistic
particles, giving high jet powers with $Q \ltsimeq Q_{\mathrm phot}$,
just as we have concluded from observational evidence in this paper.

\section{Concluding remarks}

Using two low-frequency selected radio samples with virtually complete
redshift information (3CRR and the new 7C Redshift Survey), we have
shown that:

\begin{itemize}

\item The positive correlation between the narrow emission line
luminosity $L_{\mathrm NLR}$ and extended radio luminosity $L_{\mathrm
rad}$ of FRII radio sources is intrinsic, and not an artefact of a
$L_{\mathrm NLR}$--$z$ correlation coupled with the tight $L_{\mathrm
rad}-z$ correlation present in a single flux-limited sample like 3CRR.

\item Correlations of $L_{\mathrm NLR}$ with redshift or radio
properties, such as linear size or 151 MHz rest-frame spectral index,
are either much weaker or absent.

\item The slope and normalisation of the $L_{\mathrm NLR}-L_
{\mathrm rad}$ correlation is interpreted as evidence for a close
relationship between the accretion rate, $\dot{M}$, and the jet power,
$Q$, in FRII radio sources. 

\item
The jet power $Q$ is within about an order of magnitude 
of the accretion disc luminosity $Q_{\mathrm phot}$
($0.05 \ltsimeq Q/ Q_{\mathrm phot} \ltsimeq 1$).
Values of $Q/ Q_{\mathrm phot} \sim 1$ require the 
volume filling factor $\eta$ of the synchrotron-emitting
material to be of order unity, and in addition require one
or more of the following: (i) an important contribution to the
energy budget from protons; (ii) 
a large reservoir of mildly-relativistic electrons; and (iii)
a substantial departure from the minimum energy condition in the
lobe material. 

\item Over the four orders of magnitude of radio luminosity of FRII
sources, the black hole masses lie in the range $M_{\mathrm BH} \sim
10^{7.5}- 10^{9.5} M_{\odot}$ so that there must be a change from
near-Eddington to sub-Eddington accretion as $Q$ decreases.

\end{itemize}

\section*{Acknowledgements}
We would like to thank Steve Eales, Julia Riley and David Rossitter
for important contributions to the 7C Redshift Survey. Thanks also to
Chris Simpson, Christian Kaiser, Trevor Ponman and an anonymous
referee for useful suggestions. Thanks to Robert Laing and Jasper Wall
for providing us with some 3CRR emission line fluxes ahead of
publication.  This research has made use of the NASA/IPAC
Extra-galactic Database, which is operated by the Jet Propulsion
Laboratory, Caltech, under contract with the National Aeronautics and
Space Administration. CJW thanks PPARC for support.

\end{document}